\pdfoutput=1

\documentclass[11pt]{article}

\usepackage[final]{acl}

\usepackage{times}
\usepackage{latexsym}
\usepackage{booktabs}
\usepackage{multirow}
\usepackage{graphicx}
\usepackage{xspace}
\usepackage{colortbl}
\usepackage{xcolor}         
\usepackage{amsmath}
\usepackage{subcaption}
\usepackage{tcolorbox}
\usepackage{hyperref}
\usepackage[normalem]{ulem}
\useunder{\uline}{\ul}{}
\usepackage{listings}
\lstnewenvironment{pythoncode}[1][]
{
    \lstset{
        language=Python,
        basicstyle=\small\ttfamily,
        frame=single,
        numbers=left,
        numberstyle=\tiny,
        numbersep=5pt,
        breaklines=true,
        breakatwhitespace=true,
        captionpos=b,
        showstringspaces=false,
        tabsize=4,
        xleftmargin=15pt,
        xrightmargin=15pt,
        framexleftmargin=10pt,
        framexrightmargin=10pt,
        #1 
    }
}
{}

\usepackage[T1]{fontenc}

\usepackage[utf8]{inputenc}

\usepackage{microtype}

\usepackage{inconsolata}

\title{EMO-SUPERB: An In-depth Look at Speech Emotion Recognition}

\author{Haibin Wu$^{1}$ \thanks{equal first contribution,$^\dag$equal second contribution,$^\ddagger$equal corresponding author, order is random}
, Huang-Cheng Chou$^{2\ \ast}$, Kai-Wei Chang$^{1\ \dag}$, Lucas Goncalves$^{3\ \dag}$, \\ \textbf{Jiawei Du$^{1}$, Jyh-Shing Roger Jang$^{1}$, Chi-Chun Lee$^{2\ \ddagger}$, Hung-Yi Lee$^{1\ \ddagger}$} \\
$^{1}$National Taiwan University, Taiwan \\
$^{2}$National Tsing Hua University, Taiwan \\
$^{3}$The University of Texas at Dallas, USA \\
\texttt{f07921092@ntu.edu.tw, hc.chou@gapp.nthu.edu.tw} \\
\texttt{kaiwei.chang.tw@gmail.com, goncalves@utdallas.edu} \\
\texttt{r11922185@ntu.edu.tw, jang@mirlab.org} \\
\texttt{cclee@ee.nthu.edu.tw, hungyilee@ntu.edu.tw}
}

\begin{document}
\maketitle
\setlength{\itemindent}{0em}

\begin{abstract}
Speech emotion recognition (SER) is a pivotal technology for human-computer interaction systems. 
However, 80.77\% of SER papers yield results that cannot be reproduced \citep{Antoniou_2023}.
We develop EMO-SUPERB, shorted for \textbf{EMO}tion \textbf{S}peech \textbf{U}niversal \textbf{PER}formance \textbf{B}enchmark, aims at enhancing open-source initiatives for SER.
EMO-SUPERB includes a user-friendly codebase to leverage 15 state-of-the-art speech self-supervised learning models (SSLMs) for exhaustive evaluation across six open-source SER datasets.
EMO-SUPERB streamlines result sharing via an online leaderboard, fostering collaboration within a community-driven benchmark and thereby enhancing the development of SER.
On average, 2.58\% annotations are annotated using natural language. 
SER relies on classification models and is unable to process natural languages, leading to the discarding of these valuable annotations. 
We prompt ChatGPT to mimic annotators, comprehend natural language annotations, and subsequently re-label the data. 
By utilizing labels generated by ChatGPT, we consistently achieve an average relative gain of 3.08\% across all settings.
We make all resources open-source to facilitate future developments in SER.
The source code and complete analysis are on the project website \footnote{\href{https://emosuperb.github.io/}{EMO-SUPERB Website}}

\end{abstract}

\begin{figure*}[h]
\centering
\includegraphics[width=6.0in]{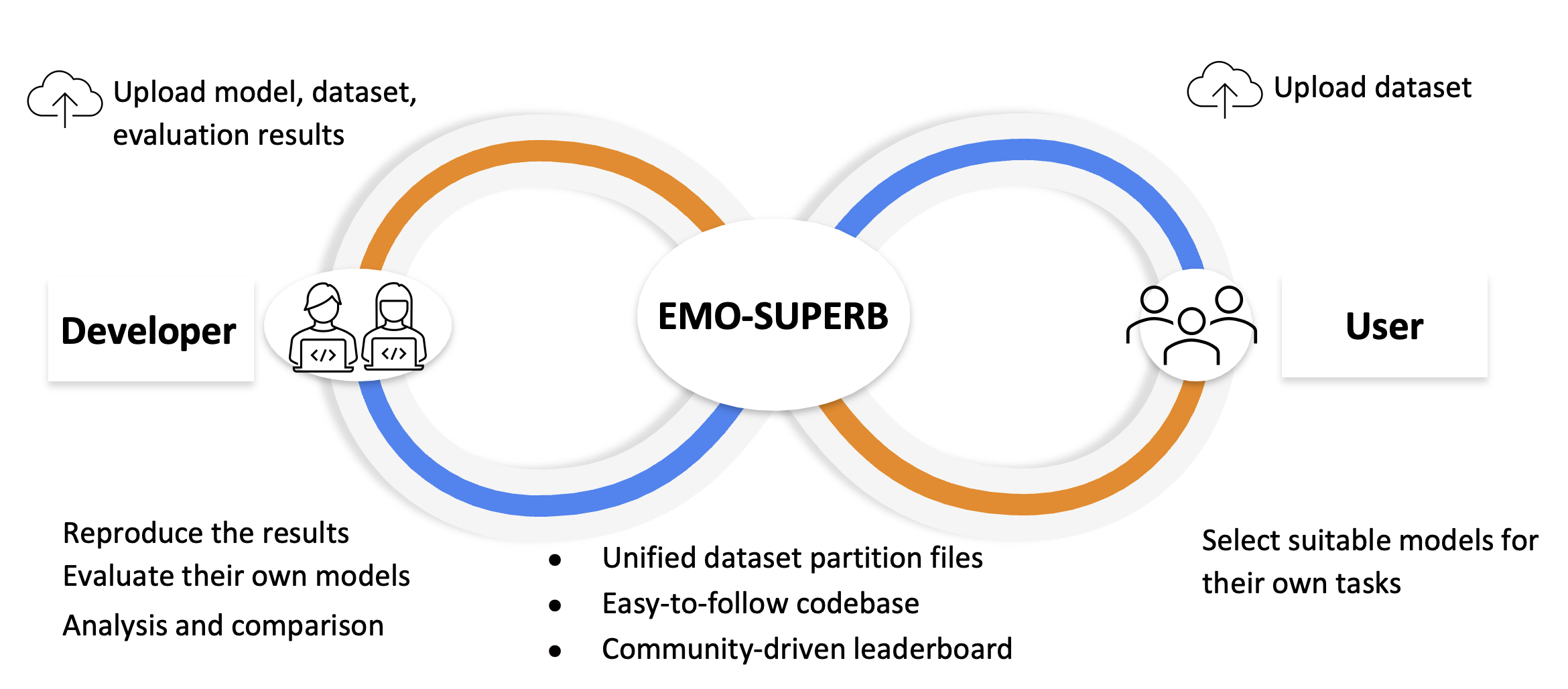}
\caption{
Demonstration for the EMO-SUPERB platform:
Developers design and evaluate SER models using our standardized dataset partition files and evaluation criteria.
Developers then contribute these prediction results to the online leaderboard, enriching the benchmark database and enabling comparative analyses with other SER models.
Finally, developers harness the visualization and statistical tools on the website to compare performance, gathering invaluable insights for future works.
From the user's standpoint, they can upload datasets and select appropriate models tailored to their individual applications.
}
\label{fig: platform}
\end{figure*}

\section{Introduction}
\label{sec: intro}

Speech Emotion Recognition (SER) aims to discern emotional cues from speech inputs, representing a pivotal technology for human-computer interaction systems. 
Recent years have witnessed significant advancements in SER.
However, there are some unsolved problems in the SER domain:

\noindent
\textbf{Issue 1}: 
Devoted annotators prefer using natural language rather than traditional emotion labels when annotating data, resulting in \textbf{typed descriptions} (e.g., ``Slightly Angry, calm'' to notify the intensity of emotion). 
While these descriptions are highly valuable, SER models, designed as classification models, cannot process natural languages and thus discard them. 
Notably, approximately 2.58\% (on average) of the annotations across all datasets use typed descriptions.

\noindent
\textbf{Issue 2}: The author of SAIL-IEMOCAP \citep{Busso_2008_5}, the most renowned SER dataset, has demonstrated that over 80.77\% of SER papers produce results that cannot be reproduced \citep{Antoniou_2023} due to the absence of released codes.

\noindent
\textbf{Issue 3}: Official data partitioning guidelines are lacking in most SER datasets. 
Consequently, different papers adopt varying partitioning strategies, leading to potential \textbf{data leakage} problems:
Typically, SER datasets comprise dialogues between two participants, denoted as Speaker A and Speaker B. 
In the process of segmenting these dialogues to isolate individual utterances, it is common to encounter scenarios where Speaker A's segments contain speech from Speaker B. 
This can cause issues because many studies adopt a straightforward approach to dividing the dataset.
They possibly allocate utterances from Speaker A for training and those from Speaker B for testing. 
However, this approach inadvertently exposes the model to Speaker B's speech during training, leading to \textbf{data leakage}. 
Studies employing this cheating partition role, with data leakage, tend to achieve 4.011\% performance improvements than those without it \citep{Antoniou_2023}. 
However, comparing settings with data leakage to those without it is unfair.

We introduce EMO-SUPERB to advance open-source initiatives in SER.
Then, we detail how to address the above three issues individually.
\begin{itemize}
    \item For \textbf{Issue 1}, we employ ChatGPT to mimic annotators, comprehend typed descriptions, and re-label the data accordingly, as shown in Section~\ref{subsec: typed description}. 
    With labels generated by ChatGPT, we consistently achieve relatively 3.08\% performance improvements across all settings.
    \item For \textbf{Issue 2}, we develop a codebase to harness 15 SSLMs, renowned for enhancing state-of-the-art performance in speech emotion recognition, for exhaustive evaluation across all open-source SER datasets in Section~\ref{subsec: SSLM-based codebase}. 
    Developers can utilize a single command line to execute both training and evaluation processes seamlessly, and we will release the easy-to-follow codebase.
    \item For \textbf{Issue 3}, we partition six open-source SER datasets and address potential data leakage issues during the partitioning process, as shown in Section~\ref{subsec: unified dataset}.
\end{itemize}

Finally, we make all datasets labeled with ChatGPT, data partition files, codes, and checkpoints open source to the community.

\section{Empower by ChatGPT }
\label{subsec: typed description}

\subsection{Importance of typed descriptions}
Emotion datasets \citep{Busso_2008_5, Chou_2017, Lotfian_2017} allow annotators to employ natural language to describe their perception of emotion corresponding to the given data if the provided label options are insufficient to capture their emotional perception fully.
These descriptions, articulated in natural language, are called typed descriptions.
Typed descriptions like ``Slightly Angry, calm'' serve to indicate the intensity of emotion, while ``Haaapy'' is used to emphasize happiness.
Appendix~\ref{Appendix: Typed Descriptions} shows more examples of the typed descriptions.

Although typed descriptions account for only approximately 2.58\% of annotations across the four emotion databases that include them, they contain valuable information for emotion perception \citep{Lotfian_2019_3, Chou_2022}. 
Because typing down the natural language to describe the emotion perception takes more time than choosing labels from label options.
Only motivated annotators will use typed descriptions.
In fact, annotators are compensated based on their working hours rather than the volume of data they handle \citep{Lotfian_2019_3}. 
Moreover, exemplary annotators receive additional bonuses. 
Some annotators invest extra time to ensure a comprehensive description of emotions to get more bonuses. 
However, SER, primarily based on classification models, cannot process natural language and consequently overlooks these valuable typed descriptions.

\begin{figure}[t!]
\centering
\includegraphics[width=3in]{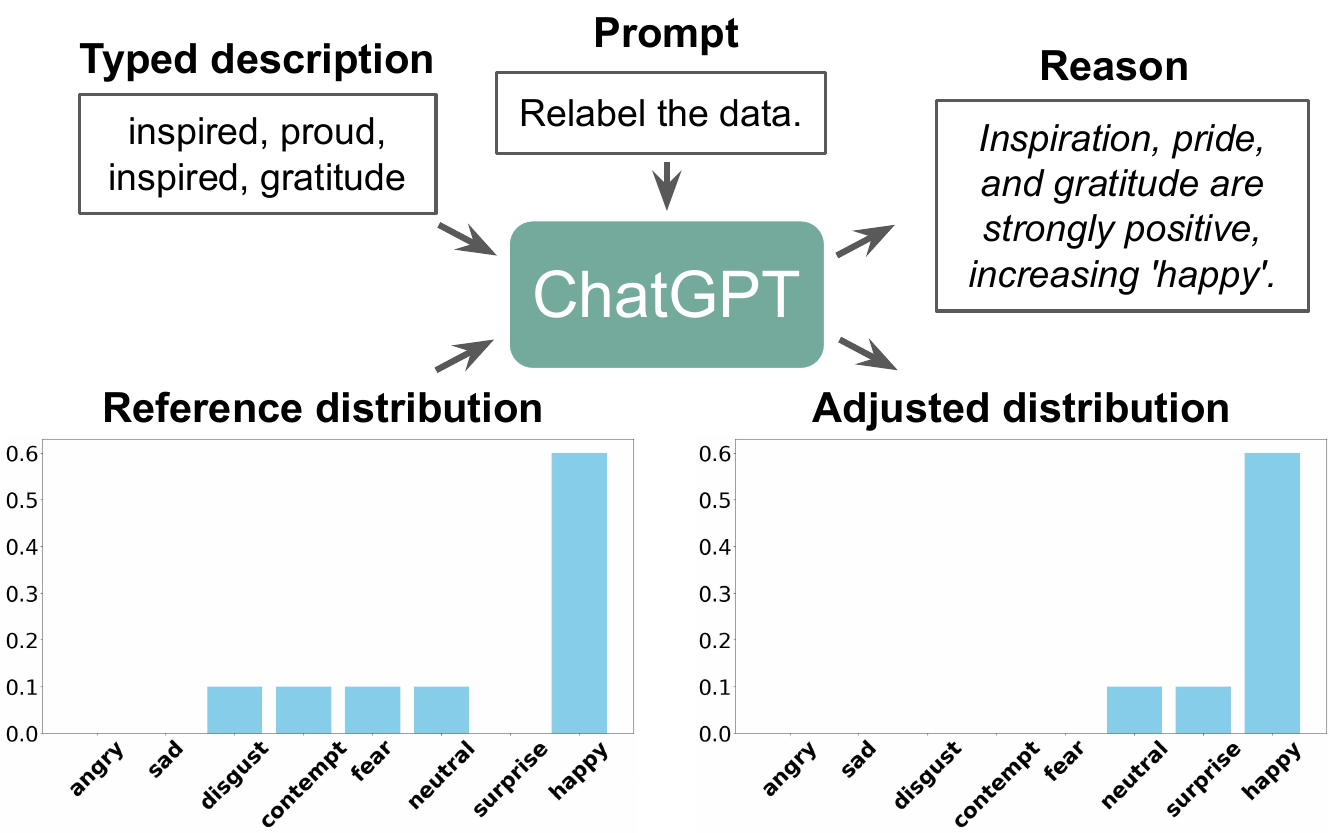}
\caption{
Labeling process using ChatGPT. Three inputs are \textbf{Typed description}, \textbf{Reference distribution}, and \textbf{Prompt}. Two outputs are \textbf{Reason} and \textbf{Adjusted distribution}. Notice that the reference distribution is calculated by the number of votes for emotion classes. In the raw annotations of an example, there are instances of disgust, contempt, fear, neutrality, and happiness (*6), resulting in values of 0.6 for happiness and 0.1 for each of the remaining appearing emotions.
}
\label{fig: GPT label}
\end{figure}

\subsection{Why using ChatGPT for relabeling}
ChatGPT \cite{Achiam_2023} exhibits a remarkable ability to comprehend and analyze natural language.
\citet{Kheiri_2023} had used ChatGPT to do effective sentiment analysis. 
Hence, we utilize ChatGPT to mimic annotators, summarizing their thoughts to re-label the data with typed descriptions.
While GPT models have been previously utilized for data labeling tasks, our approach stands out due to its innovative application in generating a distribution of labels instead of assigning a single label. 
We show that this approach leads to consistent improvements across all experimental settings as shown in Table~\ref{tab:GPTs}.


\section{EMO-SUPERB platform}
\label{sec: platform}

\subsection{Prompt ChatGPT}
\label{subsection:prompt_chatgpt}
We design a carefully crafted prompt to transform the released version of GPT-4 Turbo, a variant of ChatGPT, into a knowledgeable assistant psychologist. 
Its primary function is to generate a distribution across emotion labels based on the input typed descriptions from annotators. 

As shown in Figure~\ref{fig: GPT label}, three inputs are provided to ChatGPT: the typed descriptions, reference distributions, and a well-designed prompt.
When we prompt ChatGPT to refer to the distribution label, it fails to provide the distribution unless we supply the reference distribution.
The format of the output emotion label is also a distribution. 
Guided by the prompt, the ChatGPT can adjust or maintain the reference distribution based on the typed descriptions. 
In the prompt, we also let ChatGPT explain why it changes or doesn't change the reference distributions. 
Without this, ChatGPT might default to laziness, consistently avoiding modifying the reference distributions.
For detailed information and the final prompt, please refer to Table~\ref{tab:prompt} in Appendix~\ref{ss:prompt} due to space limitations.

We choose the MSP-PODCAST (P) dataset to verify the efficacy of our proposed prompt method in utilizing typed descriptions to improve SER, as it is the largest dataset and has the highest percentage (6.08\%) of typed descriptions among all other datasets. 
Figure~\ref{fig: GPT modified example} and the Appendix~\ref{subsubsection: output of chatgpt} show the label distributions between original and re-label ones. 
ChatGPT can understand the typed distribution and output reasonable distributions.

Ultimately, we achieved an average performance improvement of approximately 3.08\% across the 16 models on the MSP-PODCAST (P) as shown in Table~\ref{tab:GPTs} of Section~\ref{subsection:ChatGPT}.
Designing an effective prompt can enhance the accuracy of the re-labeling process.
This paper opens the door to utilizing large language models for comprehending typed descriptions. 
We welcome the community to use our user-friendly codebase to evaluate their datasets relabeled by large language models.


As shown in Figure~\ref{fig: platform}, our platform is designed to empower developers with seamless access to replicate our results, evaluate their custom SER models, compare model characteristics, and foster future SER development. 
This is facilitated by integrating three essential components: an easy-to-follow codebase, unified dataset partition files, and a community-driven leaderboard website.
Users can select SER models for their own usage.

\subsection{SSLM-based codebase}
\label{subsec: SSLM-based codebase}

\subsubsection{Framework}
Self-supervised learning (SSL) is a promising direction for developing speech models.
This approach entails training a large model with large-scale unlabeled data to obtain robust and general representations.
Notably, after pre-training, one can achieve nearly SOTA performance on downstream tasks by employing the fixed SSLMs alongside task-specific lightweight prediction heads \citep{yang2021superb}.
Furthermore, SSLMs significantly enhance SER and demonstrate SOTA performance, as evidenced in \citep{Wagner_2023}.

\begin{figure}[t!]
\centering
\includegraphics[width=3in]{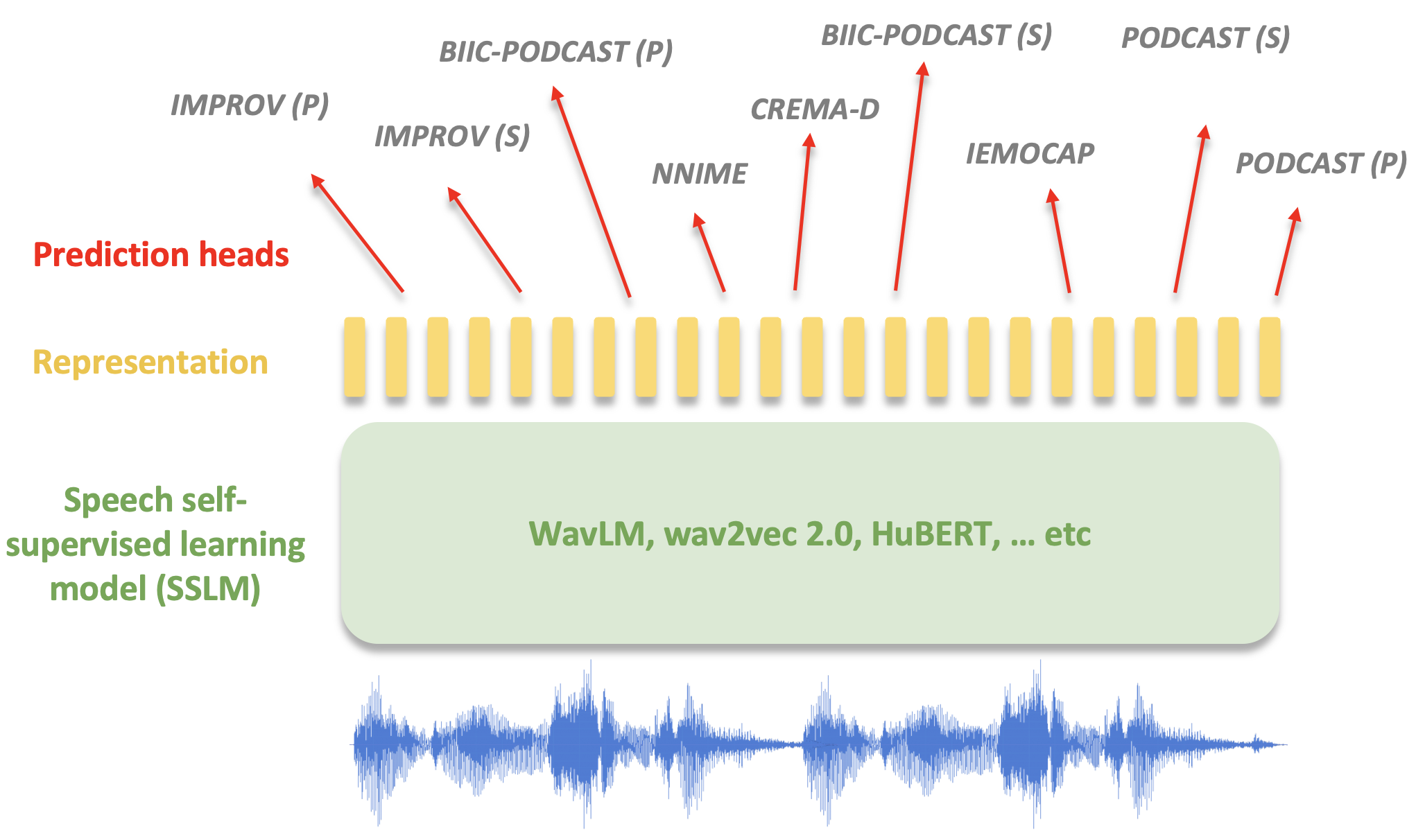}
\caption{
Illustration of SSLM-based SER
}
\label{fig: SSLM}
\end{figure}

We develop a comprehensive codebase.
The codebase depends on S3PRL \footnote{https://github.com/s3prl/s3prl} \cite{yang2021superb} to leverage 15 speech-supervised learning models as feature extractors and trains lightweight heads for exhaustive evaluation across 6 open-source SER datasets with 9 common settings, as shown in Figure~\ref{fig: SSLM}. 
The six datasets adopted are SAIL-IEMOCAP, CREMA-D \citep{Cao_2014}, MSP-IMPROV \citep{Busso_2017}, MSP-PODCAST, BIIC-NNIME \citep{Chou_2017}, and BIIC-PODCAST \citep{Upadhyay_2023}.

\subsubsection{Self-supervised learning models}

\begin{table}[b]
\centering
\resizebox{0.5\textwidth}{!}{
\begin{tabular}{lcc}
\toprule
\textbf{Model} & \textbf{Loss} \\
\midrule
Autoregressive Predictive Coding (APC) \citep{chung2019unsupervised} & Generative loss \\
VQ-APC \citep{chung2020vector} & Generative loss \\
Non-autoregressive Predictive Coding (NPC) \citep{liu2020non} & Generative loss \\
Mockingjay \citep{liu2020mockingjay}) & Generative loss \\
TERA \citep{liu2021tera} & Generative loss \\
DeCoAR 2 \cite{ling2020decoar} & Generative loss \\

WavLM \citep{chen2022wavlm} & Discriminative loss \\
Hubert \citep{hsu2021hubert} & Discriminative loss \\
wav2vec 2.0 (\textbf{W2V2}) \citep{baevski2020wav2vec} & Discriminative loss \\
Data2Vec \citep{baevski2022data2vec} & Discriminative loss \\
XLS-R \citep{babu2021xls} & Discriminative loss \\
VQ wav2vec (\textbf{VQ-W2V}) \citep{baevski2019vq} \cite{ling2020decoar} & Discriminative loss \\
wav2vec (\textbf{W2V}) \citep{schneider2019wav2vec} & Discriminative loss \\
Contrastive Predictive Coding (CPC) (\textbf{M CPC})\citep{oord2018representation}) & Discriminative loss \\
\bottomrule
\end{tabular}
}
\caption{Summary of SSLMs}
\label{table: SSLM summary}
\end{table}

We leverage two mainstream categories of SOTA SSLMs (in S3PRL), pre-trained using generative losses and discriminative losses.
We summarize them in Table~\ref{table: SSLM summary} and details can be found in Appendix~\ref{appendix:sslm} due to space limitation.

\subsubsection{Pros of the codebase}
The codebase has the following merits:
\begin{itemize}
    \item \textbf{High-performance}: Our choice to utilize SSLMs is based on their ability to consistently achieve SOTA results in speech emotion recognition, aligning with our goal to boost open-source efforts in this domain.
    \item \textbf{Affordability}: The computing barrier is greatly diminished by leveraging pre-trained SSLMs and solely fine-tuning a lightweight head, enhancing affordability for researchers from diverse backgrounds.
    \item \textbf{Reproducibility}: All codes, data partition files, and checkpoints are released, ensuring easy reproducibility of results.
    \item \textbf{Easy-to-follow}: Developers can employ a single command line to execute all training and evaluation processes, making it exceptionally user-friendly.
\end{itemize}

\subsection{Unified dataset partition rules}
\label{subsec: unified dataset}

Typically, emotion databases are collected from dialogues \citep{Busso_2008_5, Busso_2017, Chou_2017, Lotfian_2017}. 
These dialogues often involve multiple speakers engaging in intensive turn-taking, overlap, and interruption. 
The segmented utterances for each speaker commonly include speech from their conversation partners. 

For example, consider the SAIL-IEMOCAP corpus, which comprises 5 dyadic interactions (dialogues between two speakers) involving a total of ten speakers.
In 50\% of previous studies, researchers randomly divide the recordings of these ten speakers into train and test sets \citep{Antoniou_2023}.
However, due to overlap often present across speaker's segments, this practice can lead to data leakage because speaker B's speech has already been used for the model training, mentioned in section~\ref{sec: intro} (\textbf{Issue 3}).

In this study, we establish partition rules that adhere to speaker-independent criteria to mitigate the risk of leakage. 
Specifically, we ensure that all utterances from both speakers involved in dialogues are assigned to either the training or testing set. 
Further details regarding partitioning the six emotion databases can be found in Appendix \ref{sec:cv}. 
We provide the \textbf{standardization} of the training and testing splits and setups across the six public SER datasets.

\subsection{Community-driven leaderboard}
\label{subsec: Community-driven leaderboard}

The leaderboard website holds significant importance within EMO-SUPERB, continuously expanding and welcoming submissions worldwide, evolving it into a dynamic benchmark that goes beyond showcasing our own evaluation results.
To mitigate the participation barrier, the website accepts submissions with participants' own models, especially when migrating their codes to the codebase in Section~\ref{subsec: SSLM-based codebase} is not straightforward. 
Participants simply need to adhere to the data partition files outlined in Section~\ref{subsec: unified dataset}, evaluate their trained models, and submit the results.
The website also offers useful visualization (e.g. radar chart Figure~\ref{fig: radar} in Appendix~\ref{Appendix: Experiments}) and statistical tools for comparing detailed characteristics of different models, thereby enhancing future model development.

Additionally, our platform encourages community contributions of prompts and datasets with newly re-labeled typed descriptions. 
Submitters can conveniently evaluate the quality of their labeled datasets using a single command line on our codebase introduced in Section~\ref{subsec: SSLM-based codebase}.

\subsection{Artifacts}
\label{subsec: Artifacts}

Modern deep learning models present a reproducibility challenge, even with released codes, due to the potential impacts of minor hyperparameter change or package version disparities on performance. 
To assist users in debugging their training procedures, we offer Tensorboard files, hyperparameters, and pre-trained weights in our codebase. 
Furthermore, we provide downstream prediction files for several state-of-the-art models, enabling users to visualize and analyze results easily.

\section{Experimental setup}
\label{sec: experiments}

\begin{table*}[ht]
\fontsize{7}{9}\selectfont
\centering
\caption{The table summarizes the overall performance of SSLMs across the 6 public emotion datasets. \textbf{\#Par.(M)} means the number of the SSLM parameters (frozen).}
\begin{tabular}
{@{\hspace{0.2cm}}c|@{\hspace{0.1cm}}c@{\hspace{0.2cm}}c|@{\hspace{0.2cm}}c@{\hspace{0.2cm}}c@{\hspace{0.2cm}}c@{\hspace{0.2cm}}c@{\hspace{0.2cm}}c@{\hspace{0.2cm}}c@{\hspace{0.2cm}}c@{\hspace{0.2cm}}c@{\hspace{0.2cm}}c@{\hspace{0.1cm}}}
\toprule
\textbf{SSLM}       & \textbf{\#Par. (M)} & \textbf{Average} & \textbf{IMPROV (P)} & \textbf{CREMA-D} & \textbf{POD (P)} & \textbf{B-POD (P)} & \textbf{IEMOCAP} & \textbf{NNIME} & \textbf{IMPROV (S)} & \textbf{POD (S)} & \textbf{B-POD (S)} \\ \midrule
\textbf{XLS-R-1B}   & 965                 & \cellcolor{lightgray!20}{\textbf{0.38352}} & 0.552               & \cellcolor{lightgray!20}{\textbf{0.676}}   & 0.331            & \cellcolor{lightgray!20}{\textbf{0.266}}     & 0.329            & 0.209          & 0.422               & \cellcolor{lightgray!20}{\textbf{0.384}}   & \cellcolor{lightgray!20}{\textbf{0.283}}     \\
\textbf{WavLM}      & 317                 & 0.38334          & \cellcolor{lightgray!20}{\textbf{0.559}}      & 0.673            & \cellcolor{lightgray!20}{\textbf{0.350}}   & 0.252              & 0.336            & \cellcolor{lightgray!20}{\textbf{0.209}} & 0.430               & 0.369            & 0.272              \\
\textbf{Hubert}     & 317                 & 0.38331          & 0.553               & 0.675            & 0.342            & 0.262              & 0.337            & 0.197          & 0.427               & 0.383            & 0.274              \\
\textbf{W2V2   R}   & 317                 & 0.37874          & 0.555               & 0.672            & 0.331            & 0.251              & \cellcolor{lightgray!20}{\textbf{0.339}}   & 0.196          & \cellcolor{lightgray!20}{\textbf{0.433}}      & 0.363            & 0.269              \\
\textbf{Data2Vec-A} & 313                 & 0.37334          & 0.536               & 0.659            & 0.329            & 0.254              & 0.331            & 0.188          & 0.414               & 0.378            & 0.270              \\
\textbf{DeCoAR   2} & 90                  & 0.36229          & 0.512               & 0.646            & 0.308            & 0.256              & 0.320            & 0.187          & 0.405               & 0.353            & 0.274              \\
\textbf{W2V2}       & 317                 & 0.35851          & 0.469               & 0.669            & 0.321            & 0.255              & 0.306            & 0.178          & 0.396               & 0.353            & 0.281              \\
\textbf{APC}        & 4                   & 0.34975          & 0.497               & 0.608            & 0.298            & 0.249              & 0.316            & 0.186          & 0.389               & 0.340            & 0.266              \\
\textbf{VQ-APC}     & 5                   & 0.34594          & 0.497               & 0.603            & 0.296            & 0.246              & 0.312            & 0.181          & 0.389               & 0.331            & 0.259              \\
\textbf{TERA}       & 21                  & 0.34547          & 0.493               & 0.596            & 0.295            & 0.253              & 0.308            & 0.193          & 0.385               & 0.337            & 0.249              \\
\textbf{W2V}        & 33                  & 0.34212          & 0.448               & 0.612            & 0.300            & 0.246              & 0.304            & 0.188          & 0.387               & 0.336            & 0.258              \\
\textbf{Mockingjay} & 85                  & 0.33592          & 0.485               & 0.576            & 0.275            & 0.244              & 0.308            & 0.185          & 0.379               & 0.318            & 0.253              \\
\textbf{NPC}        & 19                  & 0.33150          & 0.470               & 0.570            & 0.274            & 0.240              & 0.304            & 0.172          & 0.364               & 0.333            & 0.256              \\
\textbf{VQ-W2V}     & 34                  & 0.33127          & 0.442               & 0.605            & 0.292            & 0.246              & 0.294            & 0.156          & 0.361               & 0.325            & 0.260              \\
\textbf{M   CPC}    & 2                   & 0.31508          & 0.453               & 0.529            & 0.265            & 0.228              & 0.285            & 0.175          & 0.337               & 0.318            & 0.246              \\ \midrule
\textbf{FBANK}      & 0                   & 0.19099          & 0.305               & 0.144            & 0.186            & 0.199              & 0.242            & 0.120          & 0.184               & 0.170            & 0.168              \\ \bottomrule
\end{tabular}
\label{tab:results}
\end{table*}

\subsection{Datasets}
\label{sec: setup}
We include the six public emotion datasets in the work. 
Some datasets use both primary emotions (denoted as (\textbf{P})) and secondary emotions (marked as (\textbf{S})) to allow annotators to choose single and multiple emotions, respectively.
The Appendix~\ref{appendix: dataset} presents detailed information, and Table~\ref{tab:databases} summarizes statistical data regarding the six emotion databases.
Appendix~\ref{appendix: license} outlines the license terms and usage issues.
We provide details of partitions in Appendix~\ref{sec:cv} to avoid issue 2 in Section~\ref{sec: intro}, data leakage.
The key information about these datasets is summarized as follows.
\subsubsection{The IEMOCAP}
The SAIL-IEMOCAP \citep{Busso_2008_5}, referred to as \textbf{IEMOCAP}, collects motion capture, audio, and video recordings from five dyadic conversations acted by ten professional actors in English. 
The recorded sessions were manually segmented into 10,039 utterances. 
The emotional annotations contain ten emotions and typed descriptions. 

\subsubsection{The CREMA-D}
The CREMA-D \citep{Cao_2014} contains high-quality audio-visual clips from 91 professional actors. 
There are 43 female and 48 male actors. 
There are 7,442 clips in English annotated via a crowd-sourcing platform. 
The process of perceptual annotations has three scenarios: voice-only, face-only, and audio-visual. 
In this work, we only use voice-only emotional annotations. 

\subsubsection{The IMPROV}
The MSP-IMPROV \citep{Busso_2017}, referred to as \textbf{IMPROV}, consists of high-quality audio-video sessions acted by 12 actors in English. 
All sessions are manually segmented into 8,438 clips. 
The annotation process has two scenarios: primary (\textbf{P}) and secondary (\textbf{S}) emotions. The corpus collected the typed descriptions.

\subsubsection{The POD}
The MSP-PODCAST \citep{Lotfian_2019_3}, referred to as \textbf{POD}, collected spontaneous and diverse emotional speech from various real-world podcast recordings with a commercial license. 
The labeling setting also contains primary and secondary scenarios. 
The major difference is the number of emotions in the given options. 
We use the release version 1.11 of the database, including 84,030 utterances in the train set, 19,815 in the development set, 30,647 in the test1 set, and 14,815 in the test2 set. 
We combine the test1 and test2 as the test set.

\subsubsection{The NNIME}
The BIIC-NNIME \citep{Chou_2017}, referred to as \textbf{NNIME}, consists of video, audio, and physiology recordings of dyadic conversations acted by 43 actors in Mandarin Chinese. 
All sessions are manually segmented into 5,596 clips. 
We exclude turns annotated by ``other'' from all annotators or by less than three annotators. 
The corpus also collects typed descriptions in Chinese.

\subsubsection{The B-POD}
The BIIC-PODCAST \citep{Upadhyay_2023}, referred to as \textbf{B-POD}, is a variant of MSP-PODCAST in Mandarin Chinese. 
We use the release version 1.01. 
There are 48,815 utterances in the train set, 10,845 in the development set, and 10,340 in the test set. 
At least five annotators annotate each utterance, and the emotional annotators contain primary emotions (\textbf{P}) and secondary emotions (\textbf{S}), which is the same as MSP-PODCAST.

\subsection{Preprossessing}

\subsubsection{Data Format}
We ensure the presence of audio recordings and extract them from video clips if the original datasets lack separate audio files. 
If the audio is in stereo format, we convert it to a monophonic channel. 
Furthermore, we resample the audio to 16 kHz as it is the most common sampling rate for speech processing. 
Prior to passing the speech input into modeling, we normalize it by subtracting the mean and dividing it by the standard deviation of the training set across all our experiments.

\subsubsection{Selection of emotions}
Most SER prior studies \citep{Atmaja_2022, Achiam_2023} only choose anger, happiness, sadness, and neutral state emotions as target emotions. 
In addition, they regard the excitement/joy annotations as happiness; however, excitement and happiness are not the same emotions \citep{Cowen_2017}, though those two emotions have correlations \citep{Mogilner_2011}. 

In contrast to previous approaches, we retain all original emotion labels and refrain from merging any emotions into others to balance the data (e.g., combining excitement with happiness). 
This strategy allows us to accurately assess performance and mirror natural emotion perceptions under real-world conditions.

\subsubsection{Label representation}
\label{subsubsection:label}
Inspired by \textbf{Semantics Space Theory} \citep{Cowen_2021}, we gather numerous annotations and compute a distribution-like (soft label) representation, aiming to more accurately capture the high-dimensional nature of emotion perception. Notice that these distribution-like labels are the same as the \textbf{reference distribution} used for ChaptGPT as the reference label in section~\ref{subsection:prompt_chatgpt}.

Here is one example: Let's assume we gather five annotations from five distinct raters for a single sample. 
These annotations comprise neutral (N), anger (A), anger (A), sadness (S), and sadness (S). 
Subsequently, we compute the label distributions, which in this instance are represented as (N, A, S, H) = (0.2, 0.4, 0.4, 0.0) for training SER systems. 
Additionally, in order to enhance SER performance, we employ the label smoothing technique proposed by \citep{Szegedy_2016} to refine the vector, utilizing a smoothing parameter of 0.05. 
This approach assigns a small probability to emotional classes with zero values.

\subsection{Evaluation metric}
We use the macro-F1 score \citep{opitz2019macro} to evaluate the SER performance via the Scikit-learn \citep{Fabian_2011}, considering recall and precision rates simultaneously. 
For the distribution-like multi-label training target, we select target classes by applying thresholds on the ground truth. 
A prediction is deemed successful if the proportion for a class surpasses $1/C$, where $C$ represents the number of emotional classes, aligning with the settings employed in prior research \cite{Riera_2019}. The example is in Appendix~\ref{subsection:evaluation_example}


\subsection{Objective function and training details}

Inspired by the study \citep{Cui_2019}, we adopt the class-balanced cross-entropy loss, as our primary objective function due to the imbalanced label distributions across the six databases.
Please refer to Appendix~\ref{Appendix: class-balanced cross-entropy loss} for detailed descriptions of class-balanced cross-entropy loss.
We use the AdamW optimizer \cite{Loshchilov_2019} with a 0.0001 learning rate and the batch size is 32. 
We choose the best models according to the lowest value of the class-balanced cross-entropy loss on the development set. 
We use the Nvidia Tesla v100 GPUs with 32 GB memory for all results. The total of GPU hours is around 3,300 hours. 
According to \citep{yang2021superb, tsai2022superb, feng2023superb}, SSLMs usually result in consistent results and consume large computations. 
All results in the work are single-run.
We also verify it by running experiments for small SSLMs, and the standard deviation is only less than 1\% on average.

\begin{figure}[b]
  \centering
  \begin{subfigure}[b]{0.4\textwidth}
    \centering
    \includegraphics[width=\textwidth]{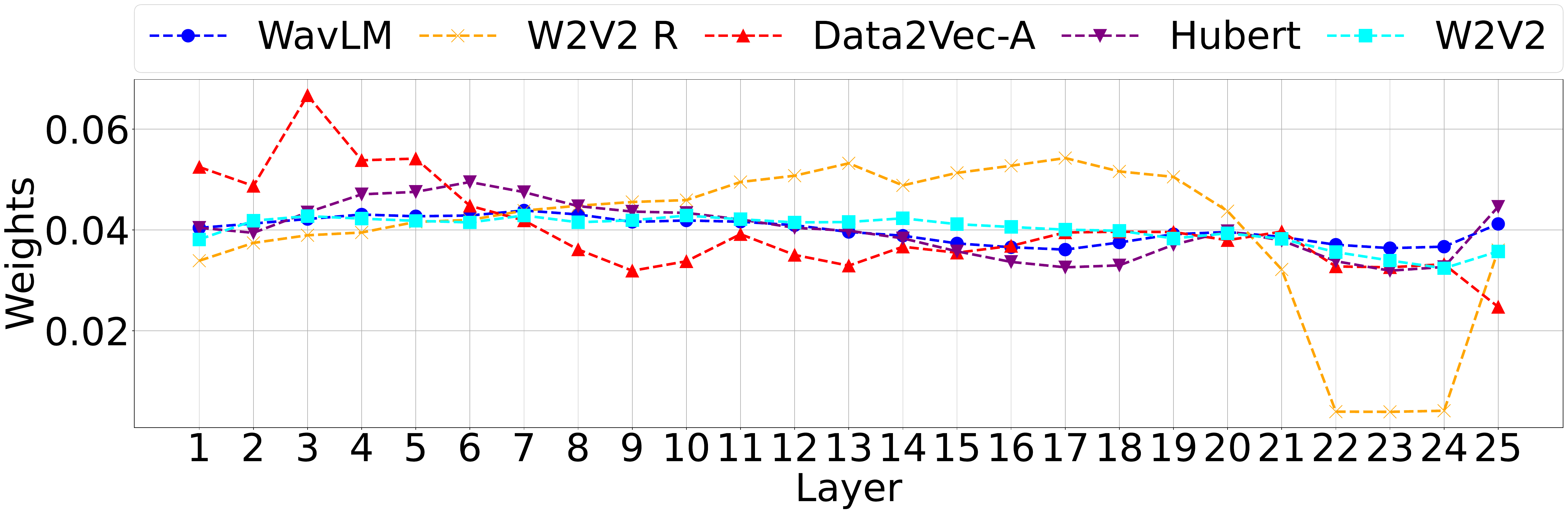}
    \caption{The layerwise weight analysis between models.}
    \label{fig:model}
  \end{subfigure}
  \hfill
  \begin{subfigure}[b]{0.4\textwidth}
    \centering
    \includegraphics[width=\textwidth]{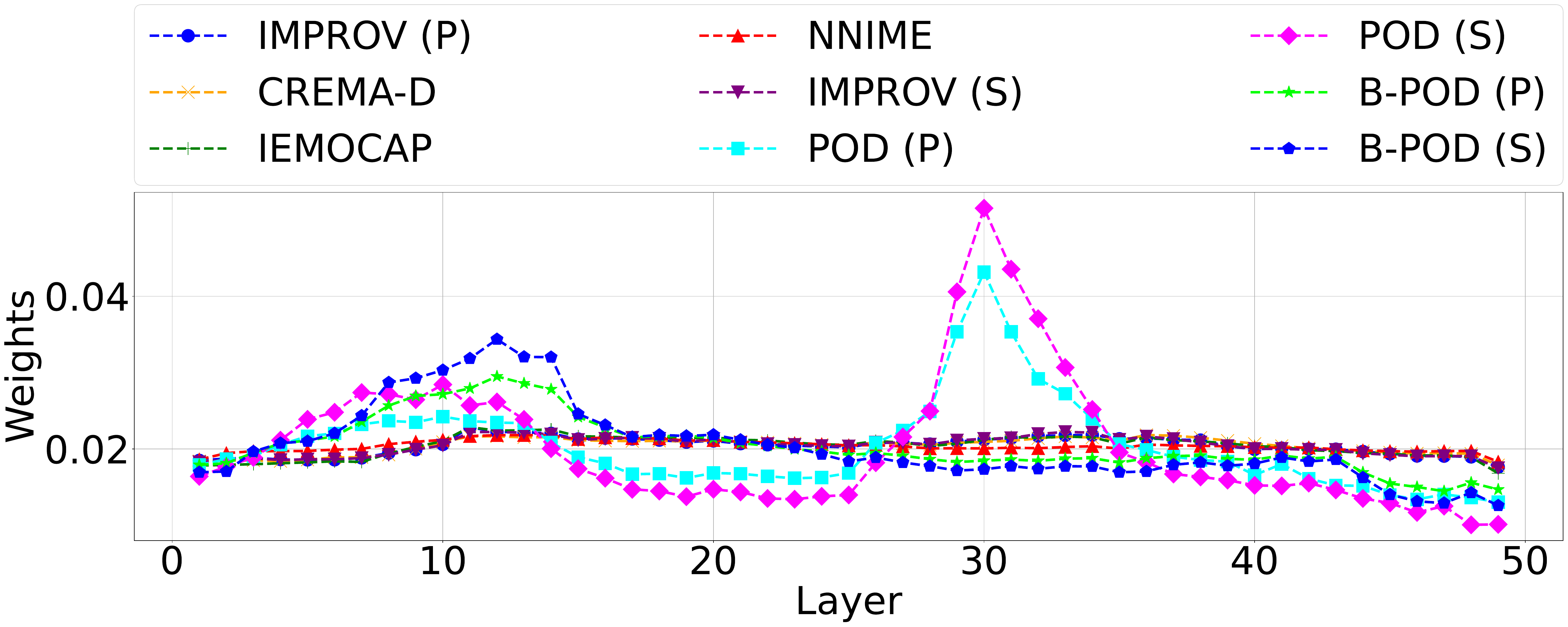}
    \caption{The layerwise weights of XLS-R-1B between datasets.}
    \label{fig:xlsr}
  \end{subfigure}
  \caption{The layerwise weights analysis.}
  \label{fig:whole}
\end{figure}

\section{Results and Analysis}
\label{sec: result}

\subsection{SSLMs for SER}
We mainly use SSLMs as our backbone models to train SER systems in the work.

\subsubsection{Overall results}
\label{subsec: overall results for SSL}
Table \ref{tab:results} summarizes macro-F1 scores obtained by 16 SSLMs and FBANK across six datasets under nine conditions. FBANK, the most commonly used speech feature, is the baseline for comparison with SSLMs. 
We have the following observations:
(1) All SSLMs exhibit significantly superior performance compared to FBANK.
Notably, XLS-R-1B achieves a remarkable improvement of relatively 100.8\% compared to FBANK.
(2) The XLS-R-1B model demonstrates the highest average performance, surpassing WavLM, which typically achieves state-of-the-art results in most speech-processing tasks. 
Despite this, WavLM still maintains considerable strength, achieving the highest performance in three out of nine conditions.
(3) Surprisingly, despite its modest 90 million model parameters, the DeCoAR 2 model outperforms the W2V2 model, which has 317 million parameters. This finding suggests that DeCoAR 2 could be an attractive choice for developers of SER facing computational resource constraints.


\subsubsection{Layer analysis}
Our training strategy involves extracting features from each layer of the SSLM, multiplying these features with layer-specific weights, and then aggregating the weighted features. These aggregated features are then fed into the downstream model. Only the layer weights and the downstream model are trainable. A large weight assigned to a specific layer suggests that the layer encodes rich emotional information. 
Additionally, we conduct a layer-wise analysis of the SSLMs. 
We select SSLMs with top-five performance, each with the same number of layers: WavLM, Hubert, W2V2 R, Data2Vec-A, and W2V2. 
We extract the layer weights from the best checkpoint of each model and normalize them using the softmax function to ensure values between 0 and 1. 
If emotion datasets contain multiple partitions (e.g., IEMOCAP and CREMA-D), we average the layer-wise weights. 
We show main results and additional layer-wise analysis can be found in Appendix~\ref{subsection:layer_appendix}.

From the model perspective (Figure \ref{fig:model}), where we sum the layer weights across all datasets for each model and plot the resulting curves. 
We have the following observations: Different models have higher weights on different layers. 
For instance, the W2V2 R has the highest weight on the 17th layer, but the Data2Vec-A's is on the third layer. 
Also, the other three models have similar patterns in emphasizing all layers. 
Additionally, it's worth noting that the weights of W2V2 R in layers 22 to 24 are considerably lower compared to those in other layers. 
We observe that the SSLMs act differently and have no clear patterns in the work.


Figure \ref{fig:xlsr} illustrates the layer weights for the state-of-the-art model, XLS-R-1B. 
Similar to other models, it exhibits a tendency to prioritize the shallow layers. 
However, there are two notable peak weights observed on the 30th layer, particularly trained on the POD (P) and POD (S) datasets. 

\begin{table}[t]
\fontsize{7}{9}\selectfont
\centering
\caption{The table presents macro-F1 scores using the integration of labels from ChatGPT.} 
\begin{tabular}{@{}c|cc|c@{}}
\toprule
\textbf{SSLM}   & \textbf{wo ChatGPT labels} & \textbf{w ChatGPT labels} & \textbf{Relative gain} \\ \midrule
\textbf{WavLM}      & 0.350             & 0.353               & 0.77\%        \\
\textbf{W2V2   R}   & 0.331             & 0.335               & 1.08\%        \\
\textbf{XLS-R-1B}   & 0.331             & 0.341               & 3.09\%        \\
\textbf{Data2Vec-A} & 0.329             & 0.338               & 2.74\%        \\
\textbf{Hubert}     & 0.342             & 0.350               & 2.22\%        \\
\textbf{W2V2}       & 0.321             & 0.325               & 1.28\%        \\
\textbf{VQ-W2V}     & 0.292             & 0.300               & 2.74\%        \\
\textbf{W2V}        & 0.301             & 0.305               & 1.58\%        \\
\textbf{CPC}      & 0.265             & 0.290               & \cellcolor{lightgray!20}{\textbf{9.45\%}}        \\
\textbf{DeCoAR   2} & 0.308             & 0.317               & 3.14\%        \\
\textbf{TERA}       & 0.295             & 0.306               & 3.52\%        \\
\textbf{Mockingjay} & 0.275             & 0.298               & 8.49\%        \\
\textbf{NPC}        & 0.275             & 0.290               & 5.75\%        \\
\textbf{VQ-APC}     & 0.296             & 0.310               & 4.94\%        \\
\textbf{APC}        & 0.298             & 0.307               & 3.19\%        \\
\textbf{FBANK}      & 0.186             & 0.186               & 0.00\%        \\ \midrule
\textbf{Average}    & 0.298             & 0.307               & 3.08\%        \\ \bottomrule
\end{tabular}
\label{tab:GPTs}
\end{table}

\subsection{Results with re-labeled data}
\label{subsection:ChatGPT}


\begin{figure}[b!]
\centering
\includegraphics[width=2.6in]{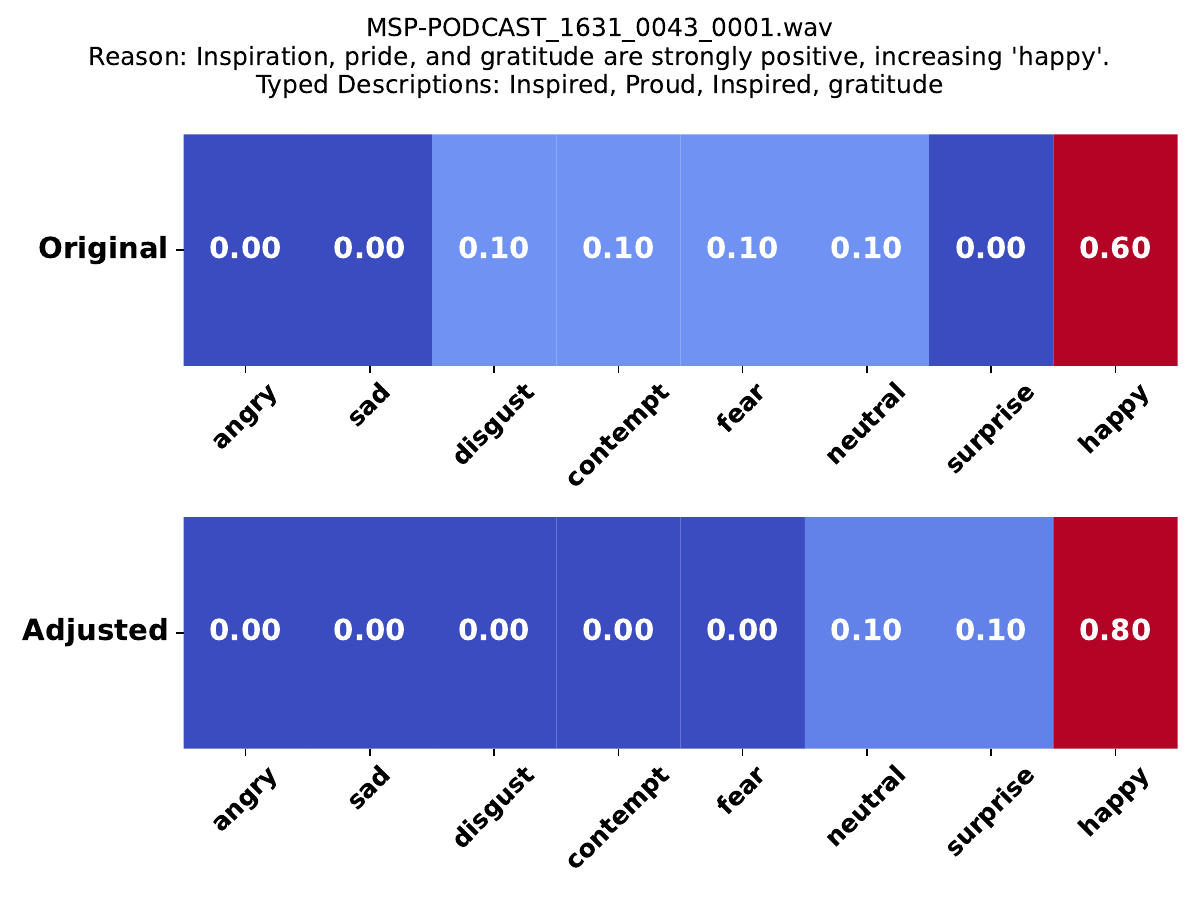}
\caption{
Original and adjusted distributions.
The original distribution, determined by tallying the votes for each emotion class, is compared with the adjusted distribution resulting from ChatGPT's re-labeling. In the raw annotations of the example, there are instances of disgust, contempt, fear, neutrality, and happiness (*6), resulting in values of 0.6 for happiness and 0.1 for each of the remaining emotions.
}
\label{fig: GPT modified example}
\end{figure}

Figure~\ref{fig: GPT modified example} shows a random selected example to compare the original distribution and relabeled distribution by ChatGPT (data sample ``PODCAST\_1631\_0043\_0001.wav'').
We observe that ChatGPT effectively comprehends the typed descriptions conveying positive emotions, thereby assigning greater weight to the ``happy'' emotion category.
More examples can be found in Figure~\ref{fig: GPT modified example - 2} to Figure~\ref{fig: GPT modified example - 5} and Table~\ref{tab:example_chatgpt} in Appendix~\ref{subsubsection: output of chatgpt}.

Table~\ref{tab:GPTs} presents the macro-F1 scores of the experiment along with the effects of incorporating data labeled by ChatGPT. 
We denote ``w/o ChatGPT labels'' and ``w/ ChatGPT labels'' to signify results without and with ChatGPT labels, respectively, while maintaining all other settings the same.
We note the following observations:
(1) The experiments involved 16 models, resulting in an average relative performance gain of 3.08\%.
(2) Particularly noteworthy is the case of CPC, which exhibits a substantial 9.45\% relative improvement.

\subsection{Takeaways}
Here are some takeaways:
(1). DeCoAR 2 achieves considerate results on SER within a modest model size, making it an appealing option for developers constrained by computation resources.
(2). SSL models trained for SER exhibit a tendency to assign greater weight to shallow layers, which are used to encode emotional information.
(3). Leveraging ChatGPT for relabeling typed descriptions holds significant promise in boosting SER performance.

\section{Ethical Considerations and Limitations}

In terms of limitations, our study solely focused on emotion datasets in English and Chinese, omitting datasets in other languages. 
Additionally, the absence of recordings featuring elderly and child speech, coupled with unknown annotator details, may hinder the representation of emotional perception across certain demographics. 
We do not address potential performance biases related to speaker gender within the SER systems. 
We only utilize ChatGPT for relabeling typed descriptions in the PODCAST dataset, given the considerable expense associated with utilizing the ChatGPT API. 
The task of designing improved prompts and labeling typed descriptions for other datasets remains for future investigation.
We haven't attempted fine-tuning the entire SSLMs due to the significant computational resources required.

\section{Conclusion and Future Work}
We propose EMO-SUPERB, an ecosystem containing user-friendly codebases, ChatGPT re-labeled datasets, pre-trained models, fair data partition files, and a community-driven leaderboard for SER. 
We effectively address open questions in SER, including (1) boosting reproducibility, (2) addressing data leakage, and (3) leveraging unused typed descriptions. 
We encourage the community to use EMO-SUPERB to develop and evaluate the SER systems. 
We plan to expand our investigation in future work by incorporating additional evaluation angles, such as calibration error and gender bias.

\newpage
\bibliography{main}

\newpage
\appendix

\begin{table}[b]
\fontsize{7}{9}\selectfont
\centering
\caption{The table summarizes license information of the six public emotion databases. The license can be accessed by clicking the word, \textbf{Agreement} or \textbf{Website}.}
\begin{tabular}{@{}ccc@{}}
\toprule
Dataset        & License   & Commercial Purposes \\ \midrule
SAIL-IEMOCAP   & \href{https://sail.usc.edu/iemocap/Data_Release_Form_IEMOCAP.pdf}{Agreement} & No                  \\
CREMA-D        & \href{https://docs.google.com/forms/d/e/1FAIpQLSdvOR994_Hsx7OkBU3oCzluXcmxw2P1nr-zBxcPgVBNLdD9Eg/viewform?usp=sf_link}{Agreement} & No                  \\
MSP-IMPROV (P) & \href{https://ecs.utdallas.edu/research/researchlabs/msp-lab/publications/AcademicLicense-MSP-IMPROV.pdf}{Agreement} & No                  \\
MSP-PODCAST    & \href{https://ecs.utdallas.edu/research/researchlabs/msp-lab/publications/Busso-FDPDTUA_V2.pdf}{Agreement} & YES                 \\
BIIC-NNIME     & \href{https://drive.google.com/file/d/1-JchUTTE0Mp2ED-gkYgMSOsJqwhN2RUF/view?usp=drive_link}{Agreement} & No                  \\
BIIC-PODCAST   & \href{http://andc.ai/}{Website} & YES                 \\ \bottomrule
\end{tabular}
\label{tab:license}
\end{table}

\begin{table*}[t]
\fontsize{7}{9}\selectfont
\centering
\caption{The table summarizes other detailed information of the 6 public emotion databases.}
\begin{tabular}{@{\hspace{0.0cm}}@{\hspace{0.05cm}}c|@{\hspace{0.05cm}}c@{\hspace{0.05cm}}c@{\hspace{0.05cm}}c@{\hspace{0.05cm}}c@{\hspace{0.05cm}}c@{\hspace{0.05cm}}c@{\hspace{0.0cm}}}
\toprule
\textbf{Database}                              & \textbf{IMPROV}  & \textbf{CREMA-D}                      & \textbf{PODCAST}     & \textbf{B-PODCAST}    & \textbf{IEMOCAP} & \textbf{NNIME}     \\ \midrule
\textbf{Citation (Paper)}                      & 323                  & \textbf{562}                          & 271                      & 1                        & \textbf{3181}         & 46                      \\
\textbf{Length (hrs)}                          & 9.53                 & 5.26                                  & 235.94                   & 147.43                   & 12.44                 & 3.26                    \\
\textbf{Length (Train, Dev., Test) (hrs)}      & (6.35,   1.59, 1.59) & (3.15,   1.05, 1.05)                  & (134.34,   31.72, 69.88) & (102.51,   22.32, 22.61) & (7.46,   2.49, 2.49)  & (1.96,   0.65, 0.65)    \\
\textbf{Sampling Rate (K Hz)}            & 44.1                 & 16                                    & 16                       & 16                       & 16                    & 16                      \\
\textbf{Max. Length (sec.)}                    & 31.91                & 5.01                                  & 11.94                    & 16.02                    & 34.14                 & 71.81                   \\
\textbf{Avg. Length (Std.) (sec.)}          & 4.09   (2.89)        & 2.54   (0.51)                         & 5.69   (2.35)            & 7.58   (3.31)            & 4.46   (3.06)         & 2.32   (2.81)           \\
\textbf{Min. Length (sec.)}                    & 0.41                 & 0.51                                  & 1.91                     & 0.51                     & 0.58                  & 0.128                   \\
\textbf{No. of Utt.}                           & 8385                 & 7442                                  & 149307                   & 70000                    & 10039                 & 5028                    \\
\textbf{Excluded Utt.}                         & 53                   & 0                                     & 2347                     & 0                        & 0                     & 568                     \\ 
\textbf{No. of Speaker}                        & 12                   & 91                                    & 2172+Unknown              & Unknown                  & 10                    & 43                      \\
\textbf{Speaker Gender}                        & 6F; 6M               & 43F; 48M                              & 904F;1268M;Unknown       & Unknown                  & 5F; 5M                & 24F; 19M            \\
\textbf{Transcriptions}                        & V                    &                                       & V                        & V                        & V                     & V                       \\
\textbf{Other Modality}                        & Video                & Video                                 &                          &                          & Video                 & Video,   Physiology     \\
\textbf{Labels/Utt.}                           & 7.36                 & 9.84                                  & 5.72                     & 3.33                     & 3.24                  & 5.12                    \\
\textbf{Raters/Utt. (Min.)}                    & 5                    & 6                                     & 5                        & 5                        & 3                     & 3                       \\
\textbf{Raters/Utt. (Mean/Std.)}               & 7.31                 & 9.84                                  & 5.72   (2.29)            & 3.33   (0.75)            & 3.24   (0.43)         & 4.97   (0.67)           \\
\textbf{Raters/Utt. (Max.)}                    & 50                   & 12                                    & 32                       & 9                        & 4                     & 6                       \\
\textbf{No. of Rators}                         & Unknown              & 2443                                  & 14363                    & Unknown                  & 12                    & 6                       \\
\textbf{Perception}                            & Observed             & Observed                              & Observed                 & Observed                 & 6 Observed, 6   Self  & Observed                \\
\textbf{Stimulus}                              & Audio-visual         & Voice-only                            & Voice-only               & Audio-visual             & Audio-visual            \\
\textbf{Emotions (P)}                    & 4                    & 6                                     & 8                        & 8                        &                       &                         \\
\textbf{Emotions (S)}                   & 10                   &                                       & 16                       & 16                       & 9                     & 11                      \\
\textbf{Setting}                               & Scripted+Improvised  & Scripted                              & Real-world               & Real-world               & Scripted+Improvised   & Improvised              \\
\textbf{Context}                               & V                    &                                       &                          &                          & V                     & V                       \\
\textbf{Speakers}                              & Actors               & Actors                                & Real-world               & Real-world               & Actors                & Actors                  \\ \bottomrule
\end{tabular}
\label{tab:databases}
\end{table*}

\section{Details of Emotion Dataset}
\label{appendix: dataset}
Table~\ref{tab:databases} summarizes six public emotion datasets, detailing the number of citations for each, the total length of audio recordings (partitioned length), original recording sampling rates, and length statistics. Additionally, it provides statistics on segmented utterances, speaker and annotator information, and collection settings for all datasets.

\subsection{License of Emotion Dataset}
\label{appendix: license}

Table \ref{tab:license} offers a concise summary of the licensing details for six emotion datasets to facilitate their accessibility within the research community. It's important to note that the majority of these datasets are restricted to academic use. However, the PODCAST and BIIC-PODCAST datasets stand out as they also offer the option for commercial licensing, albeit for a fee.

\subsection{The SAIL-IEMOCAP}
The SAIL-IEMOCAP dataset, referenced in this document as \textbf{IEMOCAP}, was meticulously assembled from the motion capture, audio, and video recordings of dyadic conversations involving ten professional English-speaking actors \citep{Busso_2008_5}. This dataset uniquely captures a blend of scripted and spontaneous dialogues, focusing primarily on scenarios that portray lovers in a relationship to elicit a rich spectrum of emotions. Each recording session featured a pair of speakers, one female and one male, engaging in interactions designed to provoke distinct emotional responses.

To ensure a diverse emotional range, the actors were provided with scripts curated to trigger specific feelings. The completed recordings were subsequently divided into 10,039 segments, each meticulously transcribed to facilitate further analysis.

Annotators then reviewed these segments, selecting emotions from a predefined list that encompassed ten distinct states: neutral, happiness, sadness, anger, surprise, fear, disgust, frustration, excitement, and an "other" category for emotions outside the listed spectrum. Additionally, a provision for typed descriptions allowed for a more nuanced annotation, accommodating the labeling of complex emotions that might not fit neatly into predefined categories. Notably, the annotation process permitted the selection of multiple emotions for a single utterance, with each segment being evaluated by at least three individuals from a pool of twelve annotators, comprised equally of actors and untrained (naive) annotators.

A significant aspect of the IEMOCAP dataset is its focus on both self-perception and observed perception annotations, differentiating it from other emotional databases. This dual approach provides a more comprehensive understanding of the emotional landscape captured within the dataset.

To address challenges related to data reproducibility, highlighted in previous research \citep{Antoniou_2023}, we have included detailed information on the dataset splits in Appendix ref{ss:cviemocap}. This is especially crucial considering the absence of standard split sets within the original corpus, a gap that our documentation aims to bridge.

\subsection{The CREMA-D}
The CREMA-D dataset, introduced by \citep{Cao_2014}, is a valuable resource comprising high-quality audio-visual clips featuring performances from 91 professional actors, including 43 females and 48 males. These actors were tasked with recording one of twelve predetermined sentences, expressing six distinct emotions: anger, disgust, fear, happiness, sadness, and a neutral state.

A notable aspect of this dataset is its extensive annotation process, involving 7,442 clips in English, evaluated by 2,443 unique annotators through a crowdsourcing platform. Each clip received feedback from at least six annotators, with each annotator attributing one of the six aforementioned emotions to the utterance. The perceptual annotation process unfolds across three distinct scenarios: voice-only, face-only, and audio-visual. In the voice-only scenario, annotators solely listen to the audio of the clips. Conversely, in the face-only scenario, annotators observe the facial expressions of the actors without accompanying audio. Finally, the audio-visual scenario allows annotators to assess both facial expressions and audio simultaneously.

For the purpose of our study on Speech Emotion Recognition (SER), we focus exclusively on emotional annotations derived from the voice-only scenario. Notably, while many previous SER investigations utilized annotations from the audio-visual scenario as a learning target, or failed to specify annotation details altogether, we opt to leverage annotations solely from the voice-only setting for our analysis. Furthermore, we provide comprehensive details regarding the dataset splits employed in our paper, as outlined in Appendix \ref{ss:cvcremad}.

\subsection{The MSP-IMPROV}

The MSP-IMPROV dataset, also known as IMPROV \citep{Busso_2017}, comprises high-quality audio-video recordings featuring interactions acted out by 12 actors in English. These sessions encompass four distinct emotions: anger, happiness, sadness, and a neutral state. Notably, each dyadic interaction involves one male and one female speaker and is recorded in four different scenarios: preparation, scripted, improvised, and improvised-scripted scenes.

To ensure comprehensive annotation, all sessions within the MSP-IMPROV dataset are manually segmented into 8,438 clips, each evaluated by at least five annotators via a crowdsourcing platform. Utilizing the quality control method proposed by \citep{Burmania_2015}, the dataset employs mechanisms to identify and eliminate unreliable annotators.

The annotation process within MSP-IMPROV presents two scenarios: primary (P) and secondary (S) emotions. In the primary scenario, annotators select one emotion from a set of five options: anger, happiness, sadness, neutral state, or "other." Secondary emotions, on the other hand, encompass a broader range, including frustration, depression, disgust, excitement, fear, and surprise. Notably, 53 utterances annotated as "other" by annotators are excluded from the dataset, although annotators have the option to provide textual descriptions when choosing this category.

Given that the dataset does not include predefined train, development, and test sets, we introduce our proposed split sets in Appendix \ref{ss:cvimprov} for the sake of clarity and consistency in subsequent analyses.

\subsection{The MSP-PODCAST}
The MSP-PODCAST dataset, or PODCAST \citep{Lotfian_2019_3}, offers a rich collection of spontaneous and diverse emotional speech extracted from real-world podcast recordings obtained under commercial licenses. Initially, the podcast recordings are segmented into individual utterances, which are then annotated via a crowdsourcing platform. Similar to MSP-IMPROV, the dataset implements quality control measures based on the methodology outlined by \citep{Burmania_2015} to ensure the reliability of annotators.

The annotation framework within MSP-PODCAST encompasses both primary (P) and secondary (S) scenarios. In the primary scenario, annotators select from a set of nine predefined emotions: anger, sadness, happiness, surprise, fear, disgust, contempt, neutral, and "other," with the option to provide additional textual descriptions if necessary. The secondary scenario expands upon the primary emotions, incorporating an additional eight classes: amusement, frustration, depression, concern, disappointment, excitement, confusion, and annoyance, totaling 17 options.

Each utterance within the dataset is evaluated by at least five unique annotators, ensuring robustness in the annotation process. The dataset version 1.11 consists of 84,030 utterances in the train set, 19,815 in the development set, 30,647 in the combined test set (test1 and test2), and 2,347 in the test3 set, which is excluded from analysis due to its private nature lacking annotations.

In total, the dataset encompasses contributions from over 2,172 distinct speakers and involves 14,363 annotators, providing a comprehensive resource for studying emotional speech.

\subsection{The BIIC-NNIME}

The BIIC-NNIME dataset, also known as NNIME \citep{Chou_2017}, is a comprehensive resource featuring video, audio, and physiology recordings of dyadic conversations acted out by 43 actors in Mandarin Chinese. These sessions are characterized by spontaneous, unscripted interactions set in everyday home environments, encompassing six emotional scenes: anger, frustration, happiness, sadness, surprise, and neutral states.

Each session within the NNIME dataset is meticulously segmented into 5,596 clips. To maintain annotation quality, utterances labeled as "other" by annotators or those annotated by fewer than three annotators are excluded from analysis. Notably, NNIME stands out from other emotion datasets due to its annotation of both speech and non-verbal behaviors, such as laughter, sighing, sobbing, and other vocal expressions.

With a total of 43 unique speakers and annotations from six different annotators, the labeling process within NNIME resembles that of the SAIL-IEMOCAP dataset. Annotators view clips sequentially and select emotions from a pool of 12 options: anger, frustration, disappointment, sadness, fear, surprise, excitement, happiness, relaxation, joy, neutral state, and "other." Moreover, annotators have the flexibility to express emotional perceptions using Chinese words.

Given the absence of standard split sets for training deep-learning models within the corpus, we provide details of our proposed split sets in Appendix \ref{ss:cvnnime} to ensure reproducibility and facilitate further research using the NNIME dataset.

\subsection{The BIIC-PODCAST}
The BIIC-PODCAST dataset, or B-PODCAST \citep{Upadhyay_2023}, presents a Mandarin Chinese variant of the MSP-PODCAST, featuring audio recordings sourced from real-world podcasts under commercial licenses. Notably, the dataset diverges from MSP-PODCAST in its labeling process, employing college students as annotators instead of utilizing a crowdsourcing platform. This approach aims to enhance quality control and ensure the reliability of annotations, a methodology similar to MSP-PODCAST's quality assessment standards.

Version 1.01 of the B-PODCAST dataset includes 48,815 utterances in the train set, 10,845 in the development set, and 10,340 in the test set. Each utterance undergoes evaluation by a minimum of five annotators. The emotional annotations within B-PODCAST encompass both primary (P) and secondary (S) emotions, mirroring the structure of MSP-PODCAST.

The primary emotions (P) include the same set of nine options as MSP-PODCAST, comprising anger, sadness, happiness, surprise, fear, disgust, contempt, neutral, and "other." Similarly, the secondary emotions (S) expand upon the primary emotions with additional classes, maintaining consistency with MSP-PODCAST.

Overall, B-PODCAST serves as a valuable resource for studying emotional speech in Mandarin Chinese, offering a curated dataset with robust annotations and quality control measures.

\section{SSLM introductions}
\label{appendix:sslm}

In our codebase, we leverage two mainstream categories of SSLMs, pre-trained using generative loss, DeCoAR 2 \cite{ling2020decoar}, Autoregressive Predictive Coding (APC) \citep{chung2019unsupervised}, VQ-APC \citep{chung2020vector}, Non-autoregressive Predictive Coding (NPC) \citep{liu2020non}, TERA \citep{liu2021tera}, and Mockingjay \citep{liu2020mockingjay}), and discriminative loss (\textbf{XLS-R-1B}) \citep{babu2021xls}, WavLM \citep{chen2022wavlm}, Hubert \citep{hsu2021hubert}, wav2vec 2.0 (\textbf{W2V2}) \citep{baevski2020wav2vec}, VQ wav2vec (\textbf{VQ-W2V}) \citep{baevski2019vq}, wav2vec (\textbf{W2V}) \citep{schneider2019wav2vec}, and Contrastive Predictive Coding (CPC) (\textbf{M CPC})\citep{oord2018representation}). 

APC employs a pretraining strategy similar to language models on a sequence of acoustic features (\textbf{FBANK}). It utilizes unidirectional RNNs to predict future FBANK frames based on past ones.
VQ-APC improves APC's representation by integrating vector-quantization (VQ) layers. 
NPC boosts APC's efficiency by replacing RNNs with CNNs for faster inference.
Mockingjay consists of Transformer encoders. It masks segments of input acoustic features along the time axis and reconstructs them during training.
TERA builds upon Mockingjay's architecture by extending the masking strategy to frequency bins. 

DeCoAR 2.0 refines Mockingjay's design by incorporating a VQ layer just before final predictions, similar to VQ-APC's approach. Its training involves larger input masks, increased batch sizes, and the utilization of more unlabeled data to improve performance.
Wav2vec introduced several architectural enhancements to refine CPC's performance.
VQ-wav2vec integrates a VQ module into wav2vec, discretizing speech into tokens post after InfoNCE pretraining. These discrete tokens are used for training a BERT model, to get contextualized representations.
Wav2vec 2.0 streamlines the vq-wav2vec pipeline into an end-to-end framework. This involves employing time masking in the latent space and substituting BERT's token prediction with InfoNCE's negative sampling.
XLS-R builds upon wav2vec 2.0, expanding its capabilities to encompass multiple languages and augmenting the dataset size.
HuBERT enables BERT's token prediction through offline clustering of representations. Predictions are made based on the clustered labels at masked locations.
WavLM, based on Hubert, introduces noise during pretraining to enhance the robustness of SSL features.

\section{Partition Setting}
\label{sec:cv}
In the speaker-independent scenario, where the model is trained on data from certain speakers and tested on data from speakers not seen during training, ensuring fair and robust evaluation is crucial. Here are the details about data partitions for experiments on the SAIL-IEMOCAP, MSP-IMPROV, CREMA-D, and BIIC-NNIME datasets.

\subsection{The IEMOCAP}
\label{ss:cviemocap}
Table \ref{tab:cviemocap} summarizes the partitioning settings for the IEMOCAP corpus. Considering each session, we define five speaker-independent splits (i.e., Dyad 1 to Dyad 5). Each session consists of two speakers engaged in dyadic interactions. In our experiments, we conduct a 5-fold cross-validation as illustrated in Table \ref{tab:cviemocap}, where each fold includes a unique combination of training, development, and test sets to ensure comprehensive evaluation of the model's performance across different dyadic interactions within the IEMOCAP corpus.

\begin{table}[h]
\centering
\fontsize{7}{9}\selectfont
\caption{IEMOCAP corpus partitions.}
\begin{tabular}{@{}cccc@{}}
\toprule
Partition & Training Set  & Development Set & Test Set \\ \midrule
1     & Dyad 1,2,3 & Dyad 4          & Dyad 5   \\
2     & Dyad 2,3,4 & Dyad 5          & Dyad 1   \\
3     & Dyad 3,4,5 & Dyad 1          & Dyad 2   \\
4     & Dyad 1,4,5 & Dyad 2          & Dyad 3   \\
5     & Dyad 1,2,4 & Dyad 3          & Dyad 4   \\ \bottomrule
\end{tabular}
\label{tab:cviemocap}
\end{table}

\subsection{The IMPROV}
\label{ss:cvimprov}
In the speaker-independent scenario, the MSP-IMPROV corpus is partitioned into six folds for cross-validation. Each fold consists of a unique combination of training, development, and test sets, as illustrated in Table \ref{tab:cvimprov}. This partitioning strategy ensures that the model is trained on interactions involving different sets of speakers and evaluated on unseen speaker combinations, facilitating robust evaluation of model generalization across various dyadic conversations within the MSP-IMPROV corpus.

\begin{table}[h]
\centering
\fontsize{7}{9}\selectfont
\caption{MSP-IMPROV corpus partitions.}
\begin{tabular}{@{}cccc@{}}
\toprule
Partition & Training Set    & Development Set & Test Set \\ \midrule
1     & Dyad 1,2,3,4 & Dyad 5          & Dyad 6   \\
2     & Dyad 1,2,3,6 & Dyad 4          & Dyad 5   \\
3     & Dyad 1,2,5,6 & Dyad 3          & Dyad 4   \\
4     & Dyad 1,4,5,6 & Dyad 2          & Dyad 3   \\
5     & Dyad 3,4,5,6 & Dyad 1          & Dyad 2   \\
6     & Dyad 2,3,4,5 & Dyad 6          & Dyad 1   \\ \bottomrule
\end{tabular}
\label{tab:cvimprov}
\end{table}

\subsection{The CREMA-D}
\label{ss:cvcremad}

In the speaker-independent scenario, the CREMA-D corpus is divided into five sets based on speaker IDs. Each set consists of a different combination of male and female speakers, as well as a distinct range of speaker IDs, as summarized in Table \ref{tab:cvcremad}. This partitioning strategy allows for a fair and balanced evaluation of models trained on CREMA-D data by ensuring that the test sets contain speakers not seen during training, thereby assessing the model's ability to generalize across different speaker characteristics and expressions. The standard partitions follow a similar methodology as the one mentioned for the IEMOCAP dataset in section~\ref{ss:cviemocap}.

\begin{table}[h]
\fontsize{7}{9}\selectfont
\caption{The CREMA-D sessions. M represents male and F represents female.}
\centering
\begin{tabular}{@{}ccr@{}}
\toprule
Session & Gender& \multicolumn{1}{c}{Speaker ID}                                                                                   \\ \midrule
1    &7M;11F& 1037-1054       \\\midrule
2    &12M;6F& 1001-1018       \\\midrule
3    &13M;6F& 1073-1091 \\\midrule
4    &9M;9F& 1055-1072       \\\midrule
5    &15M;3F& 1019-1036       \\ \bottomrule
\end{tabular}
\label{tab:cvcremad}
\end{table}

\subsection{The NNIME}
\label{ss:cvnnime}

In the speaker-independent scenario, the NNIME corpus is randomly split into five sets based on speaker IDs. Each set comprises a different combination of male and female speakers, as well as a distinct set of speaker IDs, as summarized in Table \ref{tab:cvcremad}. This partitioning strategy enables a fair and unbiased evaluation of models trained on NNIME data by ensuring that the test sets include speakers not encountered during training, thereby assessing the model's generalization capabilities across various speakers and expressions. The standard partitions follow a similar methodology as the one mentioned for the IEMOCAP dataset in section~\ref{ss:cviemocap}.

\begin{table}[h]
\fontsize{7}{9}\selectfont
\caption{The NNIME sessions. M represents male and F represents female.}
\centering
\begin{tabular}{@{}ccr@{}}
\toprule
Session & Gender& \multicolumn{1}{c}{Speaker ID}                                                                                   \\ \midrule
1    &6M;3F& 01,02,03,04,22       \\\midrule
2    &4M;4F& 05,06,07,08       \\\midrule
3    &1M;7F& 09,10,11,12 \\\midrule
4    &2M;6F& 13,14,15,16]       \\\midrule
5    &6M;4F& 17,18,19,20,21       \\ \bottomrule
\end{tabular}
\label{tab:cvnnime}
\end{table}

\begin{figure}[]
\centering
\includegraphics[width=0.6\linewidth]{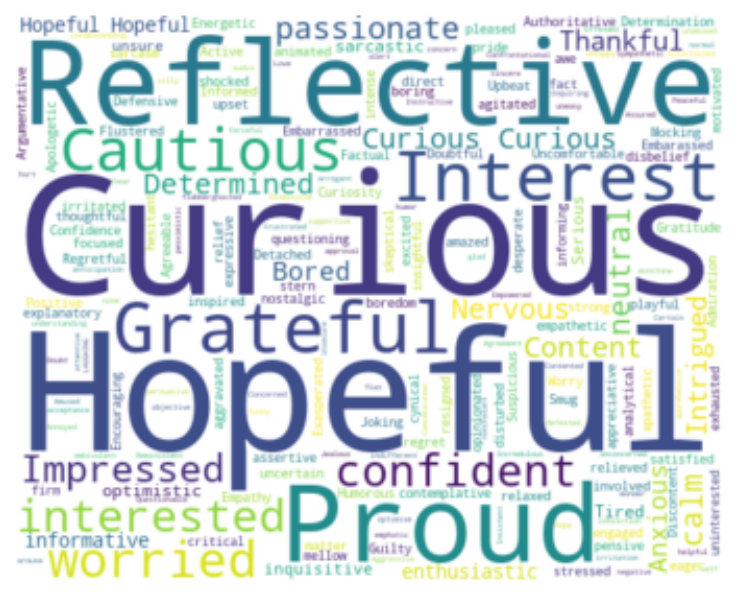}
\vspace{0.3cm}
\caption{The figures shows typed descriptions of the MSP-PODCAST for an example.} 
\label{fig:wordcloud}
\end{figure}

\section{Typed Descriptions}
\label{Appendix: Typed Descriptions}

Figure \ref{fig:wordcloud} displays a word cloud generated from typed descriptions collected in the POD emotion dataset. Word clouds are visual representations that highlight the frequency of words in a text corpus, with larger font sizes indicating higher frequencies. In this context, the word cloud provides insight into the types of words or phrases used by annotators to describe their emotional responses in the dataset.

Analyzing typed descriptions can be valuable for understanding the nuances of human emotion and improving the comprehensiveness of emotion recognition systems. By incorporating natural language processing techniques, researchers can extract valuable insights from typed descriptions to enhance the accuracy and granularity of emotion annotation in datasets like POD.


\subsection{Prompt for ChatGPT}
\label{ss:prompt}

\subsubsection{Design of Prompt}
Table~\ref{tab:prompt} shows the well-designed prompt, which contains five parts: objective, input format, example, output format, and refine and iterate. In the objective part, we clearly describe the goal of task. Then, we define the input format, including descriptions and reference emotion. Afterwards, we provide an example that provide the template the ChatGPT can follow. Finally, we ask the ChatGPT output the file in JSON format. Notice that the current version of prompt is the 14th version. In the refine and iterate part, we show the more rules that can enhance the accuracy of the output of the ChatGPT. We encourage the community to provide the designed prompt.

\subsubsection{Parameters and Cost of ChatGPT API}
Listing~\label{code:api} shows the simple code of ChatGPT API. We setup the temperature as 0. and seed as 7 in the work to keep the reproducibility. We choose the version, "gpt-4-0125-preview", as the main model. 

The average cost per data sample is \$0.0045 USD. The target emotion is represented by an 8-dimensional distribution vector. During training and development, ChatGPT examined 34.21\% and 33.32\% of the data, respectively. Out of the total 35,352 data samples available, only 12.65\% were not modified by ChatGPT. This translates to a total experimental cost of approximately \$160 USD.

\begin{pythoncode}[caption={ChatGPT API Python Code}]
response = client.chat.completions.create( 
  temperature = 0.,
  seed = 7,
  model="gpt-4-0125-preview",
  response_format={'type': 'json_object' },
  messages=[
    {
      "role": "system", 
      "content": prompt}, 
    {"role": "user", 
      "content": input_data}
  ])
\end{pythoncode}

\subsubsection{Output of ChatGPT}
\label{subsubsection: output of chatgpt}
\begin{figure}[]
\centering
\includegraphics[width=0.8\linewidth]{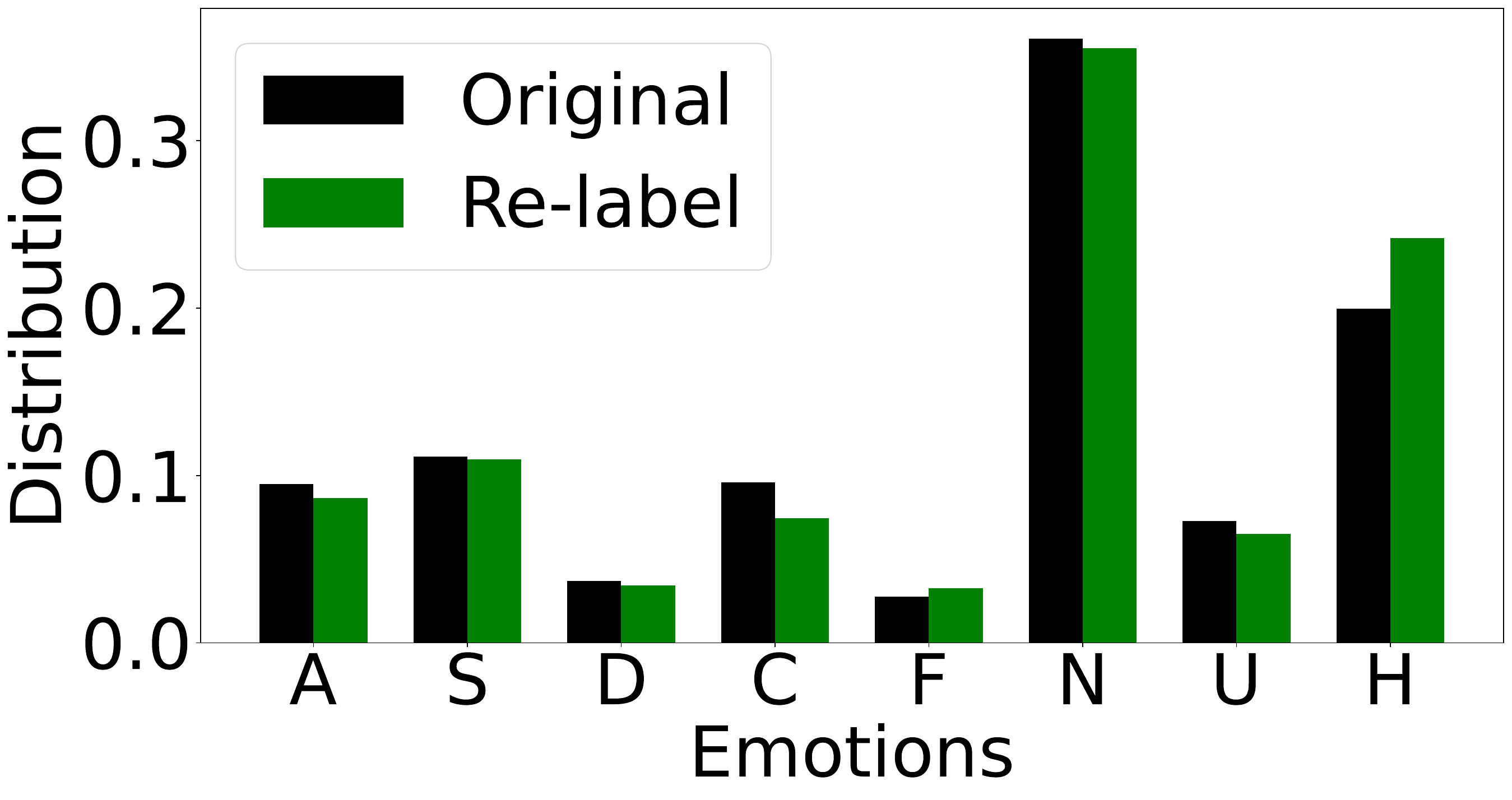}
\vspace{0.3cm}
\caption{The figures shows comparisons of original and re-labeled label distribution. Emotion includes anger (A), sadness (S), happiness (H), surprise (U), fear (F), disgust (D), contempt (C), and neutral (N).} 
\label{fig:label_distribution}
\end{figure}

\begin{figure}[t]
  \centering
  \begin{subfigure}[b]{0.45\textwidth}
    \centering
    \includegraphics[width=\textwidth]{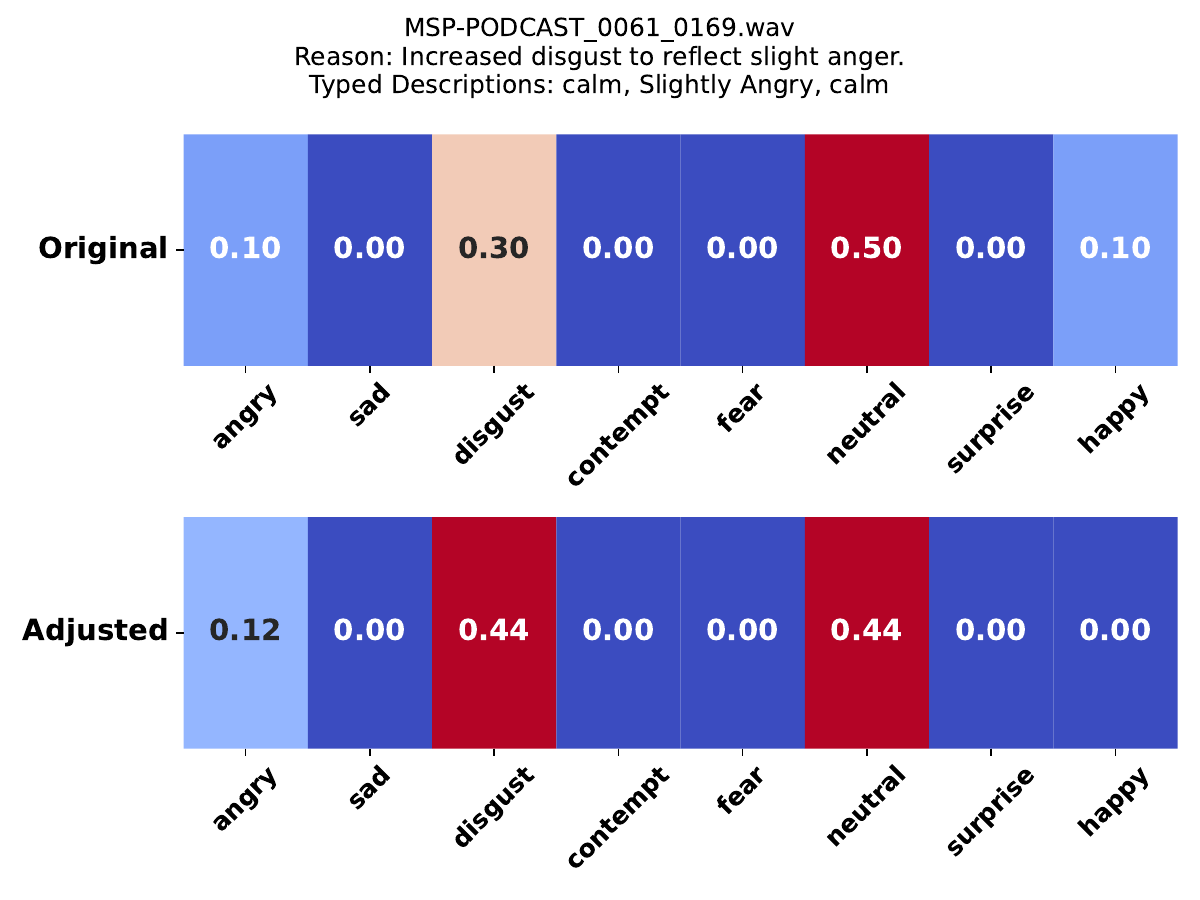}
    \caption{The second GPT modified example.}
    \label{fig: GPT modified example - 2}
  \end{subfigure}
  \hfill
  \begin{subfigure}[b]{0.45\textwidth}
    \centering
    \includegraphics[width=\textwidth]{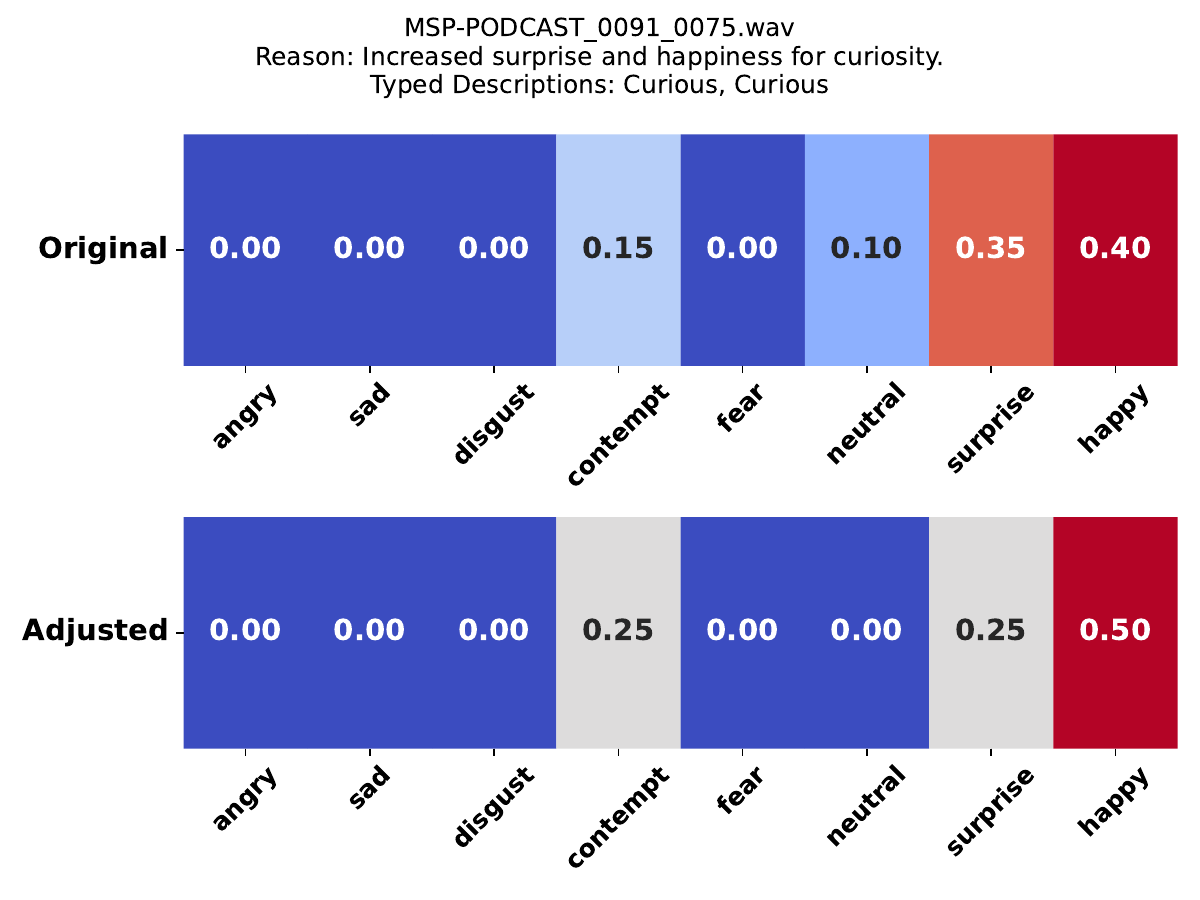}
    \caption{The third GPT modified example.}
    \label{fig: GPT modified example - 3}
  \end{subfigure}
  \hfill
  \begin{subfigure}[b]{0.45\textwidth}
    \centering
    \includegraphics[width=\textwidth]{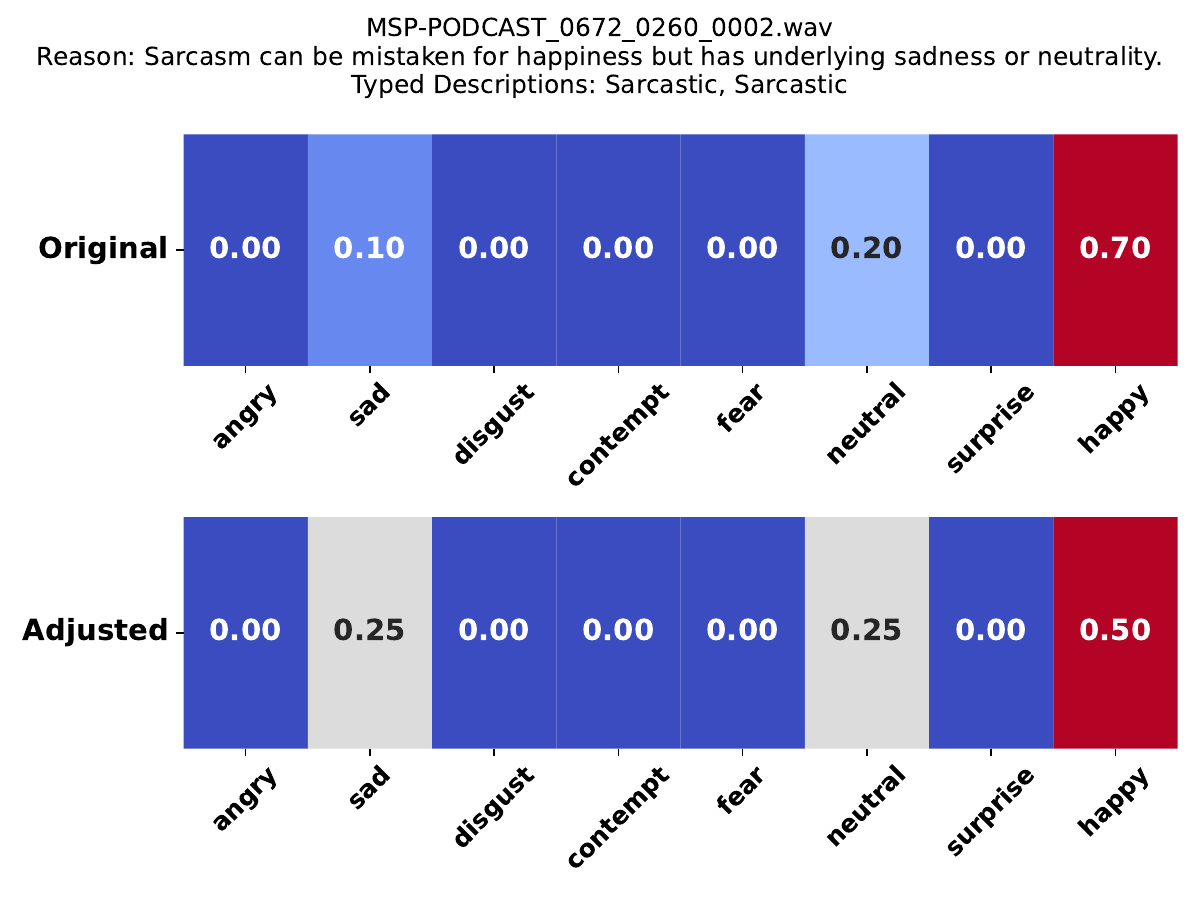}
    \caption{The fourth GPT modified example.}
    \label{fig: GPT modified example - 4}
  \end{subfigure}
  \hfill
  \begin{subfigure}[b]{0.45\textwidth}
    \centering
    \includegraphics[width=\textwidth]{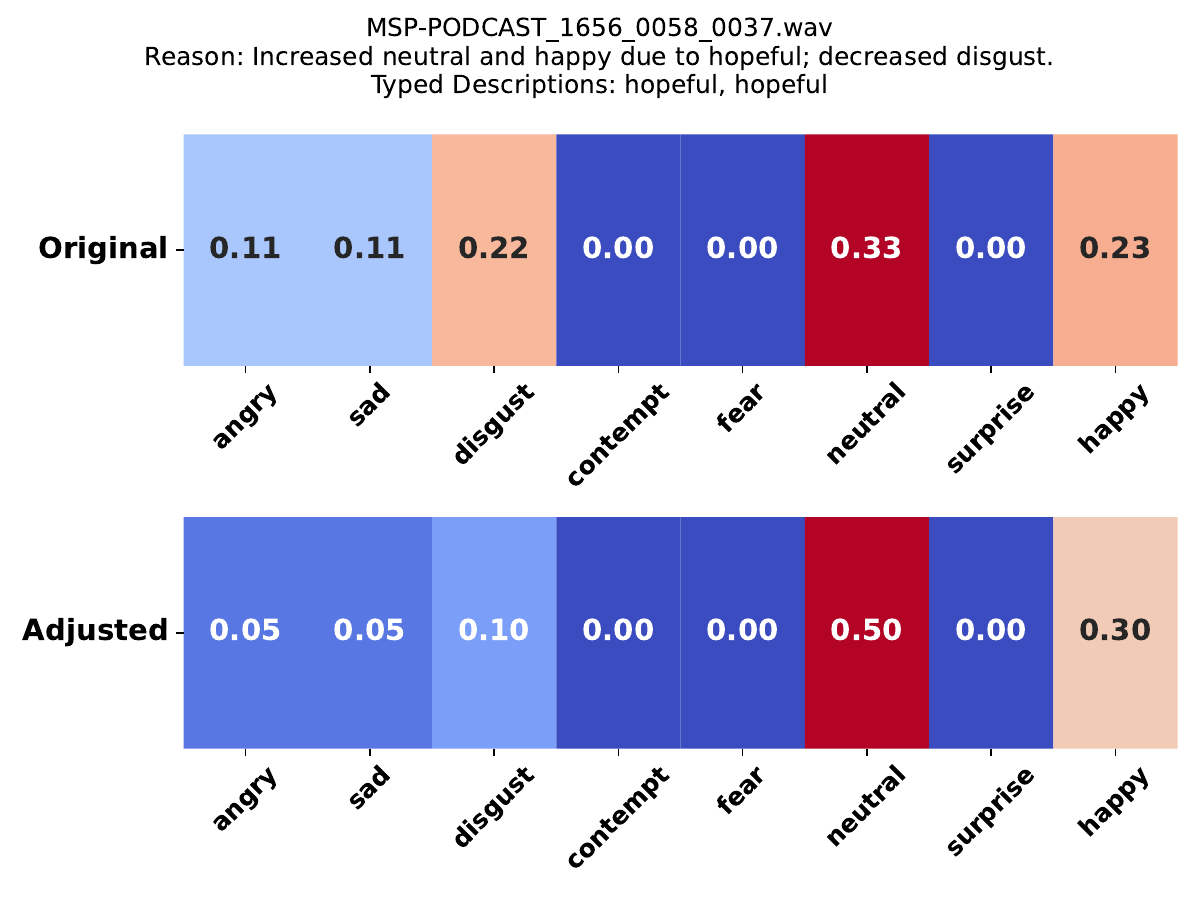}
    \caption{The fifth GPT modified example.}
    \label{fig: GPT modified example - 5}
  \end{subfigure}
  \caption{The original distribution and adjusted distribution after ChatGPT re-labeling.}
  \label{fig:GPT modified examples}
\end{figure}

Figure~\ref{fig:label_distribution} and ~\ref{fig:GPT modified examples} shows the changes of label distributions between original labels and re-label one. The ChatGPT increased the number of fear and happiness and decreased the other emotions. In addtion, the Table \ref{tab:example_chatgpt} shows ten examples including typed descriptions and reasons provided by ChatGPT.

\begin{table*}[h]
\fontsize{7}{9}\selectfont
\centering
\caption{Relabeled examples of typed descriptions with ChatGPT. ChatGPT also provides the reason to change the reference distribution or not. \textbf{Change} denotes whether the reference distribution label is changed or not.}
\begin{tabular}{@{}clll@{}}
\toprule
\multicolumn{1}{l}{Change} & Index & Typed Descriptions                & Reason                                                                            \\ \midrule 
\multirow{5}{*}{Yes}       & 01    & calm,Slightly Angry,calm          & Increased disgust to reflect slight anger.                                        \\
                           & 02    & Sarcastic ,Sarcastic              & Sarcasm can be mistaken for happiness but has underlying sadness or   neutrality. \\
                           & 03    & Curious,Curious                   & Increased surprise and happiness for curiosity.                                   \\
                           & 04    & hopeful,hopeful                   & Increased neutral and happy due to hopeful; decreased disgust.                    \\
                           & 05    & Inspired,Proud,Inspired,gratitude & Inspiration, pride, and gratitude are strongly positive, increasing   'happy'.    \\ \midrule 
\multirow{5}{*}{No}        & 06    & Tranquil                          & Maintained high neutral for tranquil's peacefulness without strong   emotions.    \\
                           & 07    & mellow,Contented                  & Kept happy at maximum for contentedness and mellowness.                           \\
                           & 08    & praising                          & Kept 'happy' at 1.0 due to 'praising' indicating strong positive   feedback.      \\
                           & 09    & Cocky,Mocking                     & Cockiness and mocking suggest contempt, but also a high degree of   happiness.    \\
                           & 10    & pride                             & Pride is a positive emotion, strongly associated with happiness.                  \\ \bottomrule
\end{tabular}
\label{tab:example_chatgpt}
\end{table*}

\begin{table*}[ht]
\caption{The Prompt for ChatGPT.}
\centering
\begin{tabular}{p{0.9\textwidth}}
\toprule
\textbf{Objective:} \\ \midrule
As a knowledgeable assistant psychologist, your role is to analyze the given words and reference labels. You generate emotion label distributions. The emotions to consider are: 'angry,' 'sad,' 'disgust,' 'contempt,' 'fear,' 'neutral,' 'surprise,' and 'happy.' The order of emotions is very important. Please provide 8-dimensional emotion distributions for these 8 emotion classes based on the user input.  \\ \midrule

\textbf{Input format:} \\ \midrule
The user input has two parts separated by \#:
The first part is the description.
The second part is 8-dimensional reference emotion distribution, 'angry,' 'sad,' 'disgust,' 'contempt,' 'fear,' 'neutral,' 'surprise,' and 'happy.' The order of reference emotion is very important.

The input has the format "descriptions\#reference emotion distribution".  Also give the reason for each data point why you want to change the reference emotion distribution.

When given the answer, you should focus 25\% on the "descriptions" and 75\% on the "reference emotion distributions". \\ \midrule

\textbf{Example:} \\ \midrule
I will give you one example: \\
User Input: Concerned,Interest\#0.0,0.0,0.0,0.0,0.0,1.0,0.0,0.0. \\
Generated Labels: \{'angry': 0.1, 'sad': 0.2, 'disgust': 0.2, 'contempt': 0.3, 'fear': 0.0, 'neutral': 0.2, 'surprise': 0.0, 'happy': 0.0, "reason": ""\} \\ \midrule

\textbf{Output format:} \\ \midrule
Reminder for the give data, It's very important to output the JSON format with index.  \\ \midrule

\textbf{Refine and Iterate :} \\ \midrule
I will give you 30 data points each time. Each data is separated by "|". It's very important.
It's very important to make sure that you complete every response for 30 data points each time.
Please reminder it. Output the JSON file that contains adjusted emotion label distributions based on reference distributions and detailed reasons why you adjust the reference emotion distributions each word by each word. It's very important.
It's very important that the JSON output file must contain the reference distributions and reasons. 
It's very important that do not contain the reference distributions and words. 
It's very important that use 15 to 20 words to explain the reason you want to change the reference distributions. 
It's very important that the sum of label distributions equals 1.
It's very important to make sure that you explain the reasons for each word in descriptions. \\ \bottomrule
\end{tabular}
\label{tab:prompt}
\end{table*}

\section{Experiments}
\label{Appendix: Experiments}
\subsection{Class-balanced cross-entropy loss}
\label{Appendix: class-balanced cross-entropy loss}
This approach proposed by the study \citep{Cui_2019} helps mitigate the impact of class imbalance by giving more weight to minority classes during the optimization process, leading to improved model performance, especially for datasets with imbalanced class distributions. The main idea is to add a weighting factor to adjust the values of the used loss function based on the inverses of the class frequency considering the training set. The factor is $ \frac{1-\beta}{1-\beta^{n_{j}}} $, where $n_{j}$ is the number of positive samples in the $j^{th}$ emotion class in the train set, and $\beta \in (0,1]$ is a hyperparameter. The number of factors to weigh the loss values equals to the number of target emotions. The CBCE value can be calculated using Eq. \ref{eq:cbl}, where $\mathcal{L_{\mathit{CE}}}^{(j)}$ is the value of cross-entropy loss for the $j^{th}$ emotion.
\begin{align} 
\mathcal{L_{\mathit{CBL}}} &=  \sum_{j=1}^{K} ( \frac{1-\beta}{1-\beta^{n_{j}}}  \cdot \mathcal{L_{\mathit{CE}}}^{(j)}). 
\label{eq:cbl}
\end{align}

\begin{figure}[b!]
\centering
\includegraphics[width=3in]{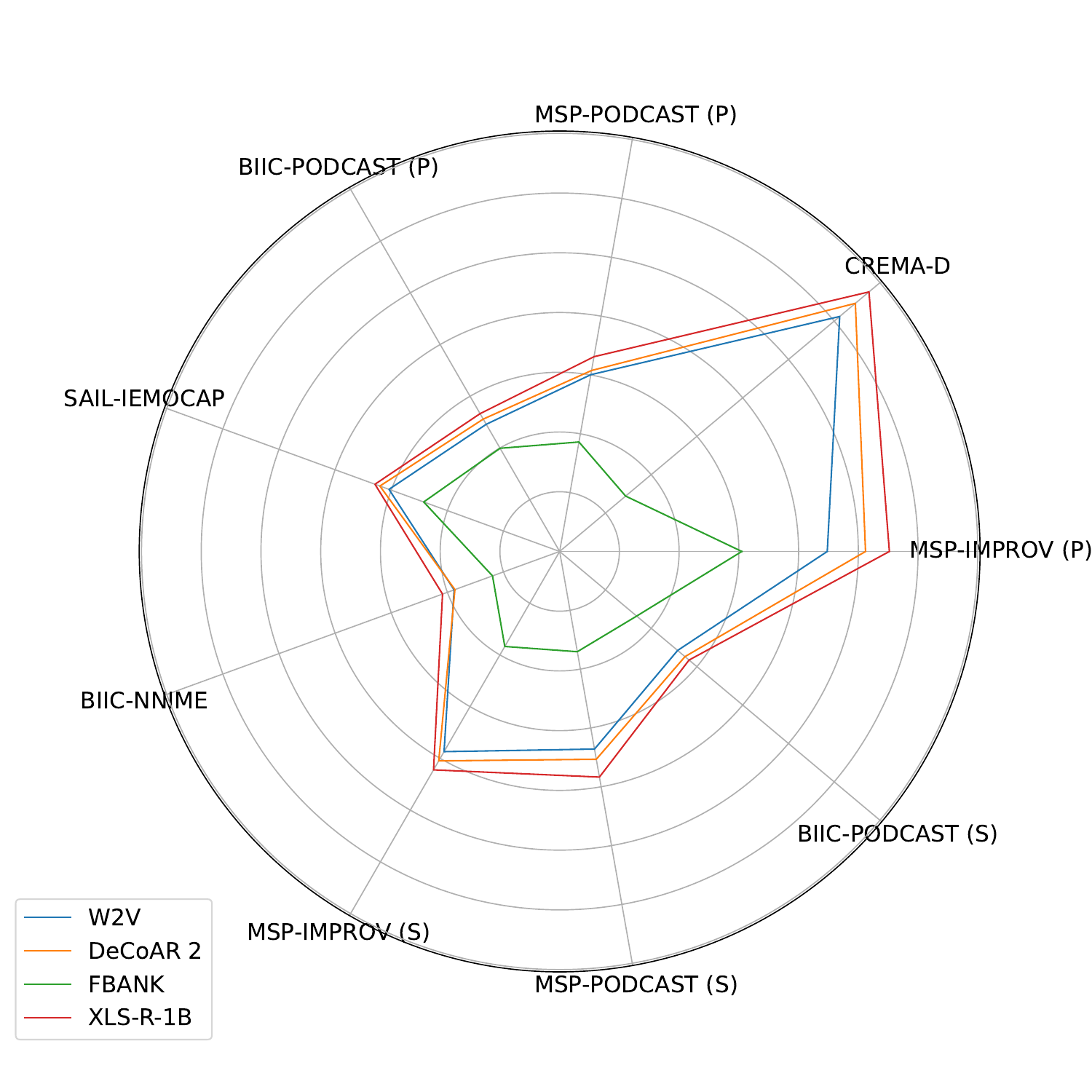}
\caption{
Demonstration of radar chart to compare four models, W2V, DeCoAR 2, XLS-R-1B and FBANK across 9 conditions.
}
\label{fig: radar}
\end{figure}

\subsection{Evaluation Example}
\label{subsection:evaluation_example}
For instance, consider a four-class emotion recognition task, and the emotion classes contain neutral, anger, sadness, and happiness. Assume we consider the predictions for three different models: (0.2,0.35,0.35,0.1), (0.1,0.45,0.45,0.0), and (0.45,0.1,0.0,0.45). The three predictions are transformed into (0,1,1,0), (0,1,1,0), and (1,0,0,1), respectively, using the threshold. In these cases, only the first two predictions are fully corrected.

\section{Analysis}
Figure~\ref{fig:five_models} and~\ref{fig:three_models} show the weights across layers.  

\subsection{Layer-wise Analysis}
\label{subsection:layer_appendix}
\begin{figure}[t]
  \centering
  \begin{subfigure}[b]{0.4\textwidth}
    \centering
    \includegraphics[width=\textwidth]{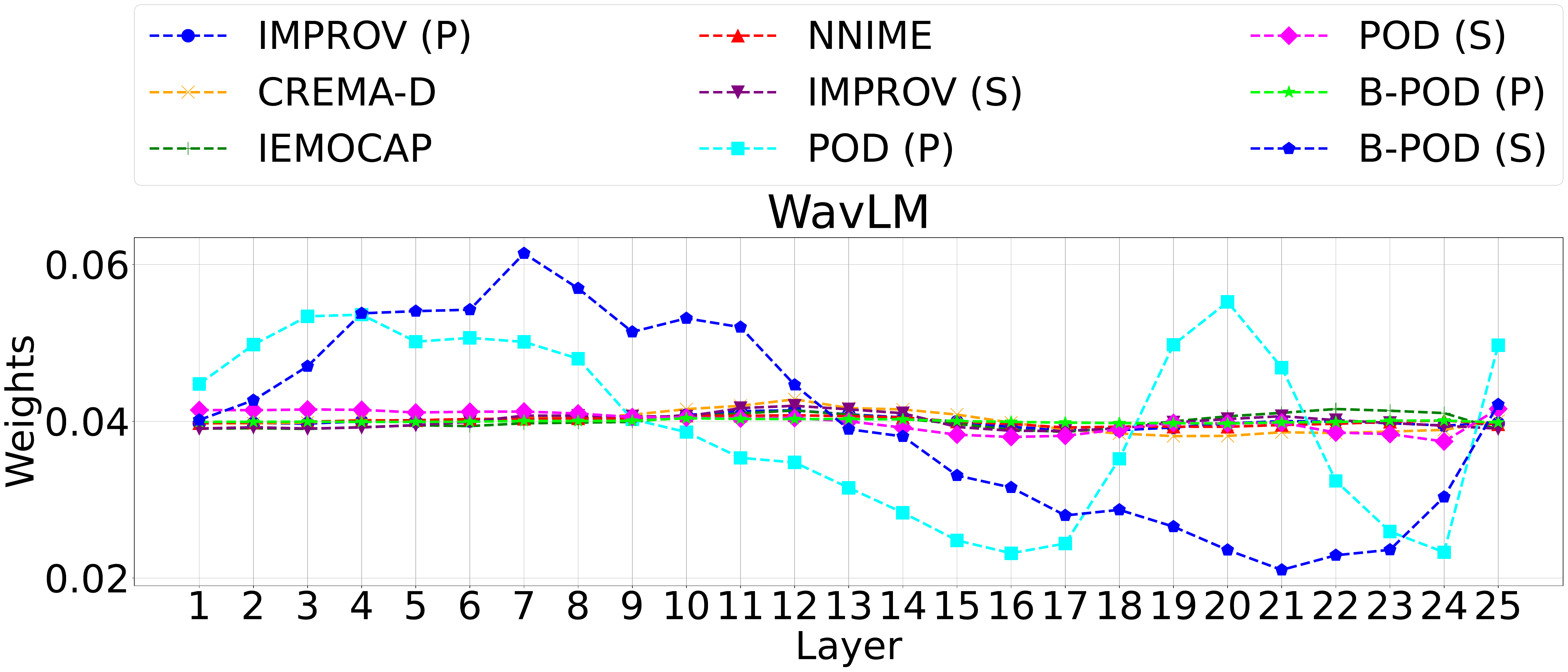}
    \caption{The layerwise weights of the WavLM.}
    \label{fig:WavLM}
  \end{subfigure}
  \hfill
  \begin{subfigure}[b]{0.4\textwidth}
    \centering
    \includegraphics[width=\textwidth]{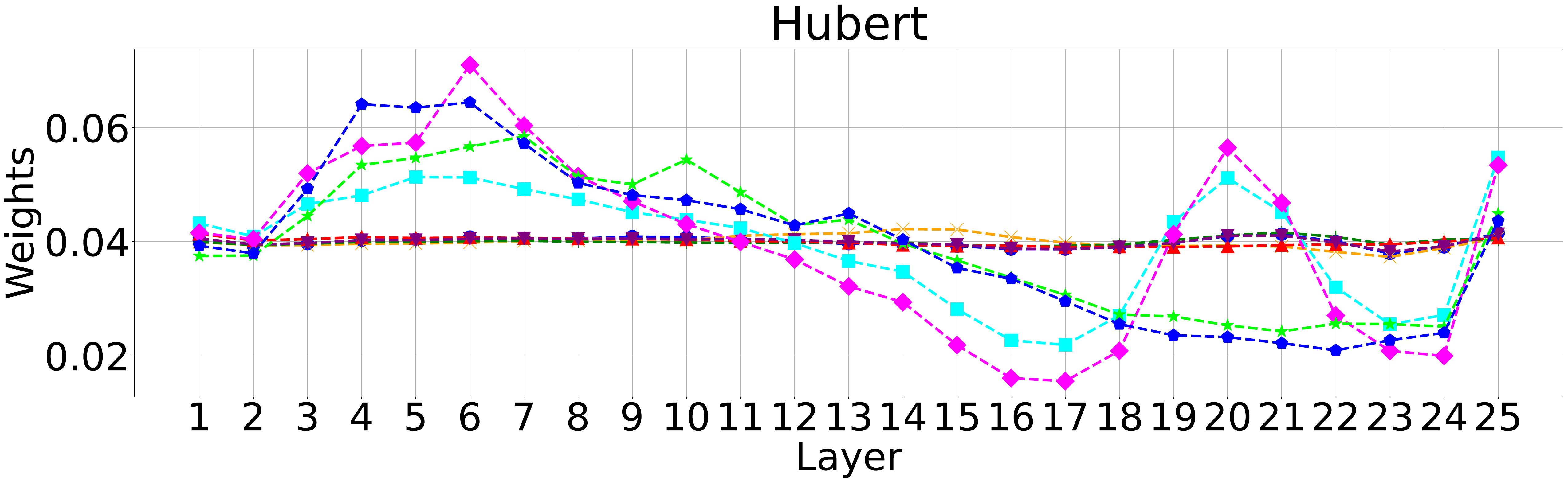}
    \caption{The layerwise weights of the Hubert.}
    \label{fig:Hubert}
  \end{subfigure}
  \hfill
  \begin{subfigure}[b]{0.4\textwidth}
    \centering
    \includegraphics[width=\textwidth]{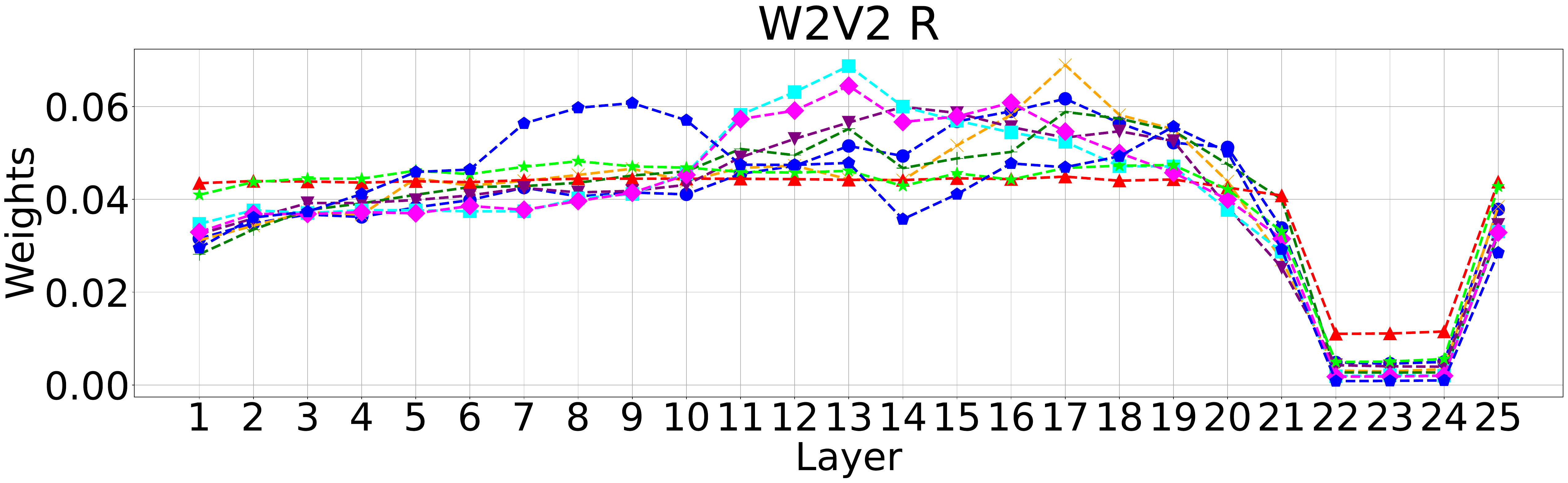}
    \caption{The layerwise weights of the W2V2 R.}
    \label{fig:W2V2R}
  \end{subfigure}
  \hfill
  \begin{subfigure}[b]{0.4\textwidth}
    \centering
    \includegraphics[width=\textwidth]{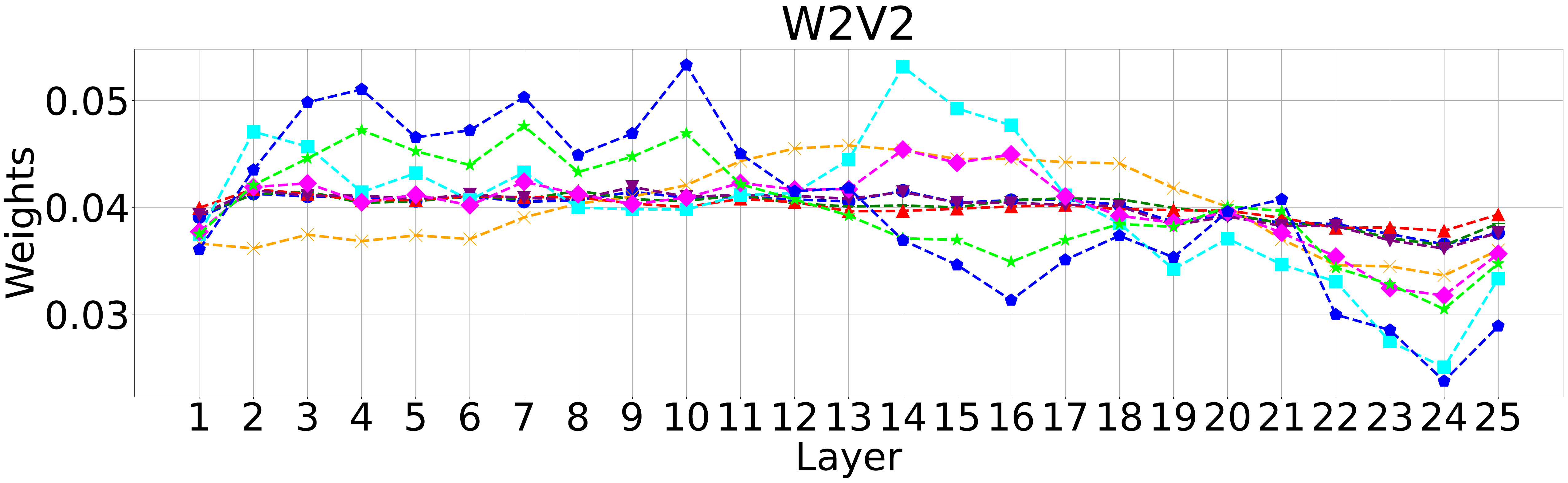}
    \caption{The layerwise weights of the W2V2.}
    \label{fig:W2V2}
  \end{subfigure}
  \hfill
  \begin{subfigure}[b]{0.4\textwidth}
    \centering
    \includegraphics[width=\textwidth]{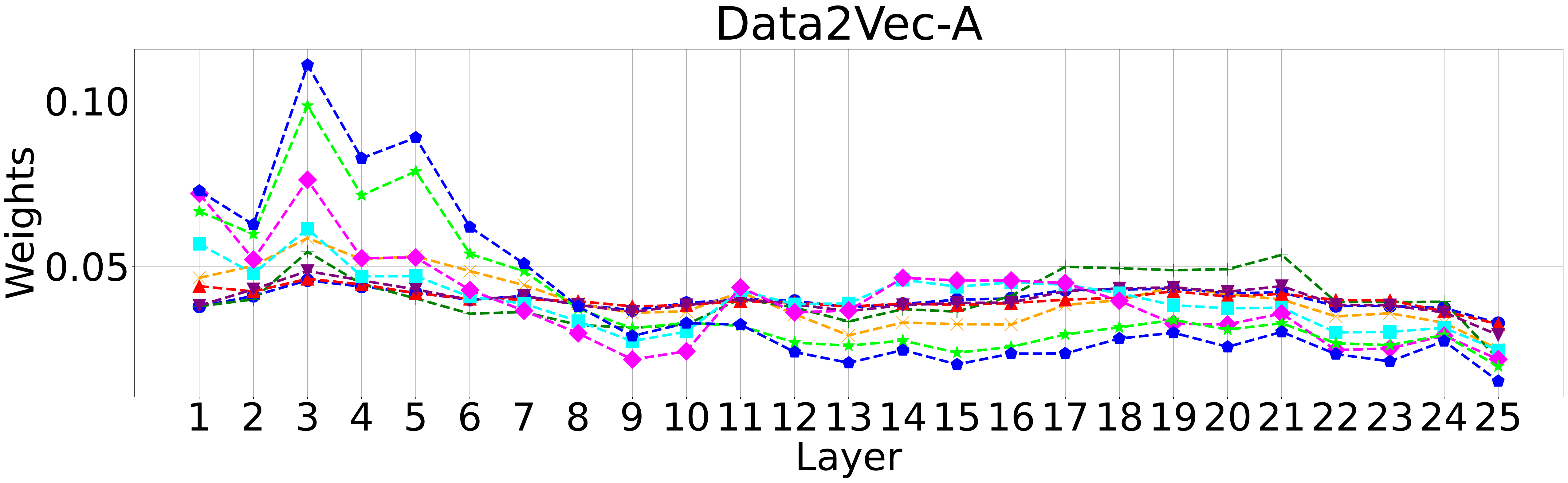}
    \caption{The layerwise weights of the Data2Vec-A.}
    \label{fig:Data2VecA}
  \end{subfigure}
  \caption{The layerwise weights analysis across 5 models.}
  \label{fig:five_models}
\end{figure}

In Figure~\ref{fig:five_models}, the patterns of POD (P) and POD (S) diverge from those of other datasets. This variance could be attributed to the incorporation of more real-life data, which likely includes additional background noise and a wider range of speakers. Concerning Chinese datasets, namely B-POD (P), B-POD (S), and NNIME, the layer weights remain relatively stable. This consistency may stem from the fact that the SSLMs were trained using English data. Across all datasets, it appears that the later layers (22nd-24th) of W2V2 are of lesser significance. Examining IEMOCAP, IMPROV (P and S), and CREMA-D, it is evident that earlier layers receive higher weighting when utilizing Data2Vec-A.

\begin{figure}[t]
  \centering
  \begin{subfigure}[b]{0.4\textwidth}
    \centering
    \includegraphics[width=\textwidth]{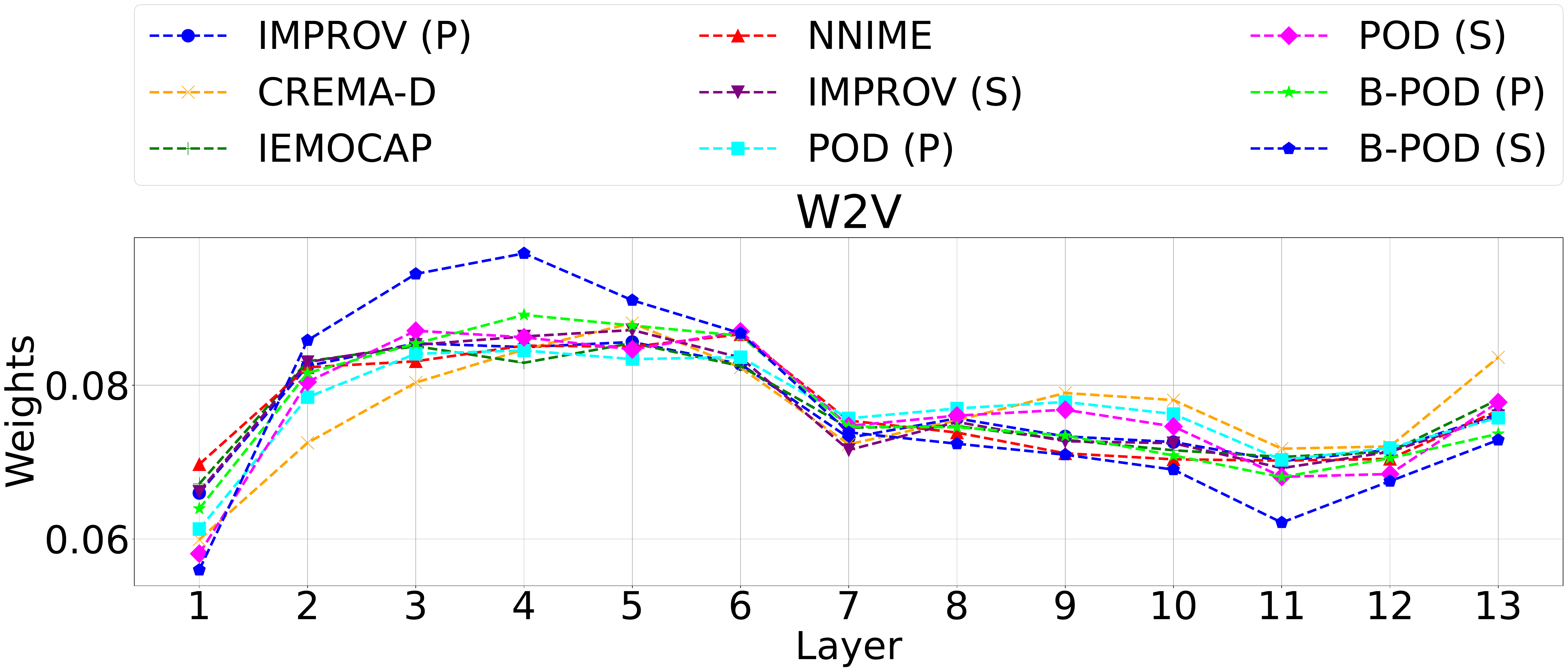}
    \caption{The layerwise weights of the W2V.}
    \label{fig:W2V}
  \end{subfigure}
  \hfill
  \begin{subfigure}[b]{0.4\textwidth}
    \centering
    \includegraphics[width=\textwidth]{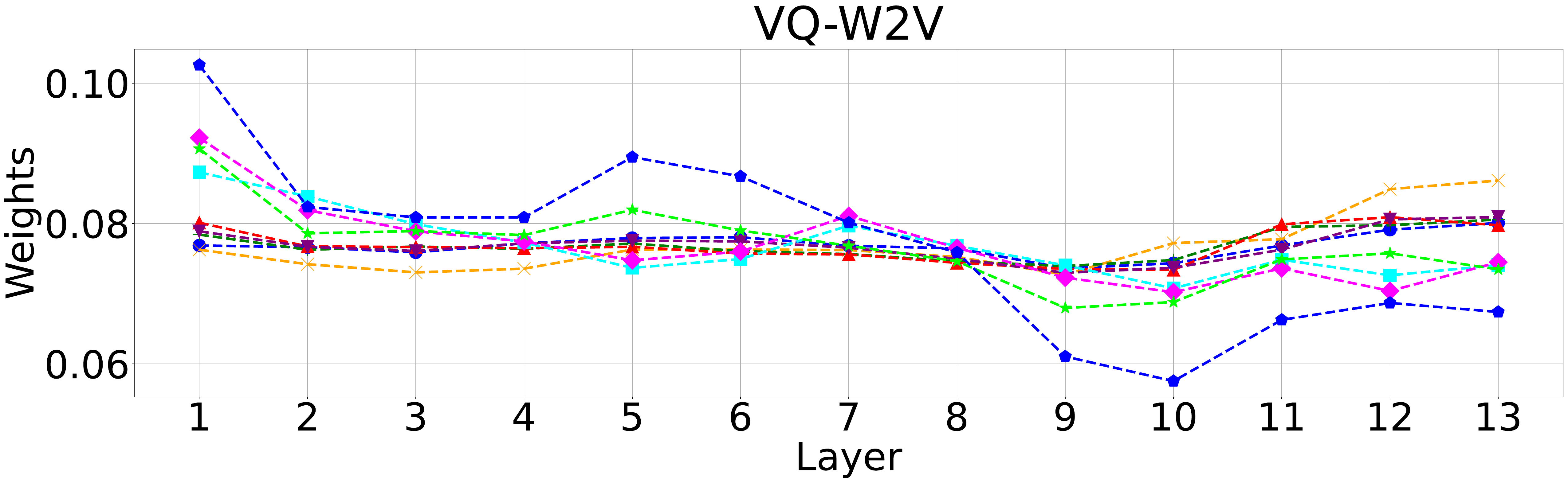}
    \caption{The layerwise weights of the VQ-W2V.}
    \label{fig:VQW2V}
  \end{subfigure}
  \hfill
  \begin{subfigure}[b]{0.4\textwidth}
    \centering
    \includegraphics[width=\textwidth]{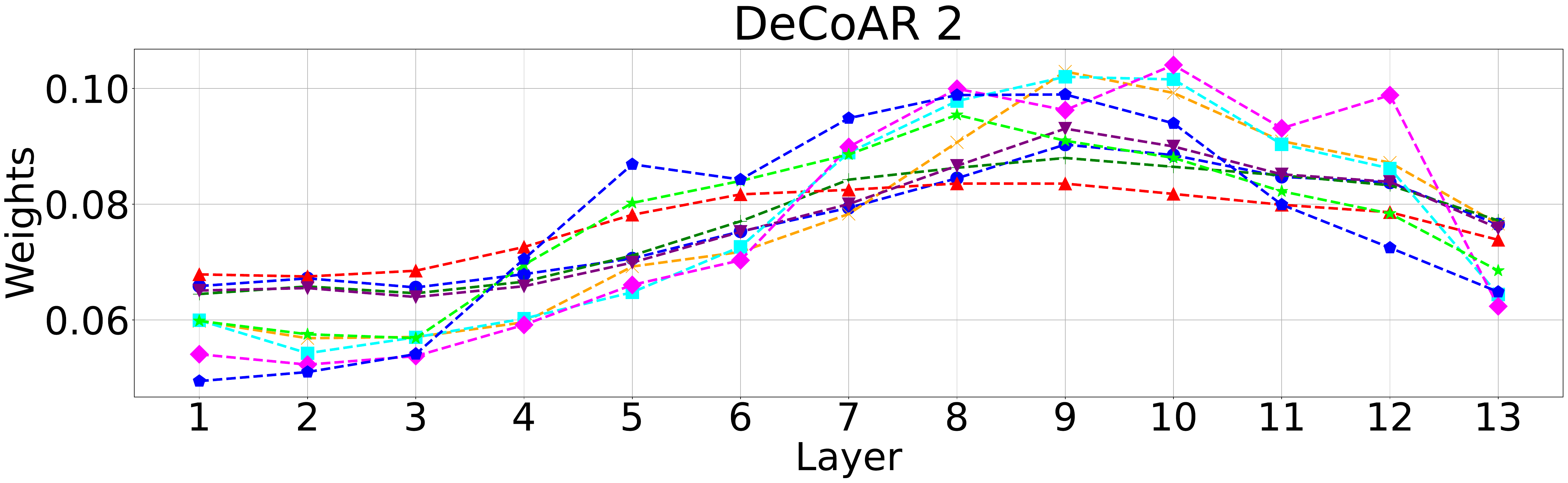}
    \caption{The layerwise weights of the DeCoAR 2.}
    \label{fig:DeCoAR2}
  \end{subfigure}
  \caption{The layerwise weights analysis across three models.}
  \label{fig:three_models}
\end{figure}

In Figure~\ref{fig:three_models}, our observations align with those reported in the study by \citet{Li_2023}, indicating a concentration of the model's focus on the shallow layers (1st-5th). However, it is noteworthy that \citet{Li_2023} identified the sixth layer as the one providing the most effective representation for the SER task. Additionally, the DeCOAR model relies on the later layers for the SER task.

\end{document}